\documentclass[fleqn]{aa}
\usepackage{amssymb}
\usepackage{graphicx}
\usepackage{pdflscape}
\usepackage{grffile}
\usepackage{natbib}
\usepackage{txfonts}
\usepackage{color}
\usepackage{multirow}
\usepackage{multicol}
\usepackage{tabu}
\usepackage{caption}
\usepackage{subcaption}
\usepackage{epsfig}
\usepackage{chngcntr}
\usepackage{ulem}

\newcommand{\Ref}[1]{(\ref{#1})}
\newcommand{\vect}[1]{\boldsymbol{#1}}

\newcommand{\Ave}[1]{\left\langle{#1}\right\rangle}
\newcommand{\ave}[1]{\bigl\langle{#1}\bigr\rangle}

\newcommand{\eb}[1]{\left({#1}\right)}

\newcommand{\ee}{\mathrm{e}}
\newcommand{\ii}{\mathrm{i}}

\newcommand{\Msolar}{\ensuremath{\mathrm{M}_\odot}}

\newcommand{\vtheta}{\vect{\theta}}

\newcommand{\deltagal}{\delta_{\mathrm{g}}}
\newcommand{\kappagal}{\kappa_{\mathrm{g}}}

\newcommand{\deltam}{\delta_{\mathrm{m}}}

\renewcommand{\d}[0]{{\rm d}}
\newcommand{\e}[0]{{\rm e}}
\renewcommand{\i}[0]{{\rm i}}
\renewcommand{\vec}[1]{\vect{#1}}

\newcommand{\pprime}[0]{{\prime\prime}}
\newcommand{\msol}[0]{{\rm M}_\odot}
\newcommand{\ita}[1]{\textit{#1}}

\begin{document}
\title{A comparison of the excess mass around CFHTLenS galaxy-pairs to
  predictions from a semi-analytic model using galaxy-galaxy-galaxy
  lensing}

\author{P. Simon\inst{1}, H. Saghiha\inst{1}, S. Hilbert\inst{2,3},
  P. Schneider\inst{1}, C. Boever\inst{1}, A. H. Wright\inst{1}}
\titlerunning{Excess mass around CFHTLenS galaxy-pairs compared to a
  SAM} \authorrunning{Simon et al.}

\institute{
$^1$ Argelander-Institut f{\"u}r Astronomie, Universit{\"a}t Bonn, Auf dem H{\"u}gel
71, 53121 Bonn, Germany\\
$^2$ Exzellenzcluster Universe, Boltzmannstr. 2, 85748 Garching, Germany\\
$^3$ Ludwig-Maximiliams-Universit\"at, Universit\"ats-Sternwarte, Scheinerstr. 1, 81679 M\"unchen, Germany
}

\date{Received / Accepted }

\abstract{The matter environment of galaxies is connected to the
  physics of galaxy formation and evolution. In particular, the
  average matter distribution around galaxy pairs is a strong test for
  galaxy models. Utilising galaxy-galaxy-galaxy lensing as a direct
  probe, we map out the distribution of correlated surface
  mass-density around galaxy pairs in the Canada-France-Hawaii
  Telescope Lensing Survey (CFHTLenS). We compare, for the first time,
  these so-called excess mass maps to predictions provided by a recent
  semi-analytic model, which is implanted within the dark-matter
  Millennium Simulation. We analyse galaxies with stellar masses
  between \mbox{$10^9-10^{11}\,\msol$} in two photometric redshift
  bins, for lens redshifts \mbox{$z\lesssim0.6$}.  The projected
  separation of the galaxy pairs ranges between
  \mbox{$170-300\,h^{-1}\,\rm kpc$}, thereby focusing on pairs inside
  groups and clusters. To allow us a better interpretation of the
  maps, we discuss the impact of chance pairs, i.e., galaxy pairs that
  appear close to each other in projection only. We introduce an
  alternative correlation map that is less affected by projection
  effects but has a lower signal-to-noise ratio. Our tests with
  synthetic data demonstrate that the patterns observed in both types
  of maps are essentially produced by correlated pairs which are close
  in redshift ($\Delta z\lesssim5\times10^{-3}$). We also verify the
  excellent accuracy of the map estimators. In an application to the
  galaxy samples in the CFHTLenS, we obtain a $3\sigma-6\sigma$
  significant detection of the excess mass and an overall good
  agreement with the galaxy model predictions. There are, however, a
  few localised spots in the maps where the observational data
  disagrees with the model predictions on a \mbox{$\approx3.5\sigma$}
  confidence level.  Although we have no strong indications for
  systematic errors in the maps, this disagreement may be related to
  the residual B-mode pattern observed in the average of all
  maps. Alternatively, misaligned galaxy pairs inside dark matter
  halos or lensing by a misaligned distribution of the intra-cluster
  gas might also cause the unanticipated bulge in the distribution of
  the excess mass between lens pairs.}


\maketitle

\section{Introduction}
\label{sect:introduction}

According to the standard paradigm of galaxy physics, there must exist
a strong correlation between the distributions of dark matter and
galaxies \citep{2010gfe..book.....M}. Semi-analytic models (SAMs) of
galaxies are one method utilised to simulate the various complexities
of galaxy physics by combining analytic prescriptions, where possible,
with results from numerical simulations of the cosmological
dark-matter density field (e.g., \citealt{2005Natur.435..629S} and
\citealt{2014arXiv1410.0365H}, H15 hereafter). However, while the main
physical mechanisms of galaxy evolution seem to be identified by
observations, specific details concerning feedback and baryonic
physics (of, in particular, star formation) are somewhat unclear.  As
a result, prescriptions that are encoded in SAMs are frequently
phenomenological and employ free parameters calibrated to observables.
Typical examples include the stellar-mass function of galaxies or, as
for H15, the fraction of quiescent galaxies as function of redshift.

The galaxy-matter correlations as function of galaxy type, cosmic time
(redshift), spatial scale, and environment encode useful statistical
information about galaxy physics, and they can be used to test
SAMs. An important tool for gathering this information is that of weak
gravitational lensing, an effect which shears the shapes of distant
galaxies (``sources'') through differential light deflection in the
presence of a tidal gravitational field between the source and the
observer \citep[for a review,
see][]{schneider2006gravitational}. Importantly, the shear distortion,
as described by the theory of general relativity, does not depend on
the nature of the gravitating matter, making lensing an ideal probe
for dark-matter physics. 

Galaxy-galaxy lensing is one application in weak lensing that
considers the correlation between positions of galaxies (lenses) and
the shear signal of source galaxies in the background. This is a probe
of the radial profile of the surface mass-density of matter around an
average lens
\citep[e.g.,][]{2017MNRAS.465.4204C,2015MNRAS.452.3529V,2012ApJ...759..101C,2006MNRAS.368..715M}.
Galaxy-galaxy-galaxy lensing is a recent extension of the second-order
galaxy-galaxy lensing that probes the third-order correlations between
the projected matter density and the galaxy number-density with two
correlation functions
\citep{2013MNRAS.430.2476S,2008A&A...479..655S,2005IAUS..225..243W,2005A&A...432..783S}. For
the scope of this study, we only work with the lens-lens-shear
correlation function and use this correlator synonymously with
galaxy-galaxy-galaxy lensing.  This kind of statistic is similar to
galaxy-galaxy lensing in the sense that it measures the mean
tangential shear around pairs of lenses or, after application of a
lensing mass-reconstruction, the lensing convergence that is
correlated with lens pairs \citep{2012A&A...548A.102S}. Because the
lensing convergence is essentially the projected matter density on the
sky, the map produced from a galaxy-galaxy-galaxy lensing visualises
the typical matter environment of galaxy pairs in projection.  With
galaxy-galaxy-galaxy lensing being a connected three-point correlation
function (by definition), the map shows the convergence in excess of
the convergence around two individual galaxies. We therefore refer to
this map as `excess mass map'. Since the introduction of
galaxy-galaxy-galaxy lensing, alternative lensing measures of mass
around average lens pairs have also been proposed and obtained from
data, partly to probe the filamentary structure of the cosmic web
\citep{2017arXiv170208485E,2016MNRAS.457.2391C,2006MNRAS.367.1222J}.

As discussed in \citet{2012A&A...547A..77S} and recently shown by
\citet[][hereafter S17]{2016arXiv160808629S} galaxy-galaxy-galaxy
lensing can test SAMs by comparing model predictions for the average
matter density around galaxy pairs to measurements. The analysis in
S17 is based on the CFHTLenS\footnote{\url{http://cfhtlens.org/}}
measurements in \cite{2013MNRAS.430.2476S}, S13 herafter, and uses
lens galaxies with stellar masses between
\mbox{$\sim10^9-10^{11}\,{\rm M}_\odot$} and redshifts below
\mbox{$z\lesssim0.6$}. S17 find the H15 model to be in good agreement
with the CFHTLenS observations, while other models strongly disagree
with the data. In addition, S17 argue that, for the same data,
galaxy-galaxy-galaxy lensing has more discriminating power in this
test than second-order galaxy-galaxy lensing. In particular, the
strong dependence on galaxy morphology and galaxy colour makes
galaxy-galaxy-galaxy lensing a powerful test for galaxy
models.

We revisit the CFHTLenS data and the most promising galaxy model in
S17 (i.e., the H15 model) for this paper, in an effort to gain more
insight into the matter-galaxy relation on spatial scales of a few 100
$h^{-1}\,\rm kpc$.  In particular, we create mass maps such that we
are able to probe the matter environment of galaxy pairs on smaller
physical scales than the related aperture statistics that are utilised
in S17.  These maps offer a better intuitive interpretation of the
signal, by directly mapping out the average surface-matter density
that is correlated with lens pairs for a fixed separation of the
pair. Conversely the aperture statistics are useful for quantitative
measurements because they are an average of the noisy
galaxy-galaxy-galaxy-lensing correlation function for a broad range of
separations, closely connected to the angular galaxy-matter bispectrum
on the sky. We perform measurements of the excess mass around CFHTLenS
galaxy pairs and, in a first study of this kind, compare these maps to
the H15 model predictions. Moreover, for future studies, we introduce
and investigate a promising new variant of the excess mass which we
designate the `pair convergence'. In tests with synthetic data and
from theoretical arguments, we find that it is less affected by
nonphysical lens pairs that are merely close on the sky in projection
(referred to as `chance pairs' in the following).

The structure of the paper is as follows. In
Sect. \ref{sect:formalism}, we introduce our notation as well as the
definitions of all correlation functions relevant for second-order and
third-order galaxy-galaxy lensing. Section \ref{sect:ExcessMass}
introduces the excess mass map and establishes its relation to the
correlation function of galaxy-galaxy-galaxy lensing.  Importantly, we
discuss the effect of chance pairs on the excess map, and we introduce
the pair convergence, a variation of the excess mass that is less
affected by chance pairs. Our two investigated data sets, the
simulated mocks and CFHTLenS data, are briefly described in
Sect. \ref{sect:data}.  In Sect. \ref{sect:methods} we outline the two
estimators for the excess mass (or the pair convergence). We apply
these estimators to CFHTLenS data and simulated H15 data in
Sect. \ref{sect:results}, verify our computer implementation of the
CFHTLenS mapping code, and quantify its accuracy.  Finally, in
Sect. \ref{sect:discussion} we discuss our conclusions for the excess
mass around CFHTLenS pairs and how they qualitatively compare to the
H15 predictions.

\section{Formalism}
\label{sect:formalism}

\subsection{Lensing notation}

Let $\deltam(\vec{x})$ be the fractional density contrast
$\delta\rho_{\rm m}/\bar{\rho}_{\rm m}$ in the matter-density field
$\rho_{\rm m}(\vec{x})$ relative to the mean density
$\bar{\rho}_{\rm m}$ at a comoving position $\vec{x}$. We define
positions $\vec{x}$ with respect to a fiducial light ray with the
observer at the origin \citep{2001PhR...340..291B}. For this, let
$f_K(\chi)\,\vec{\theta}$ at comoving distance $\chi$ be the
transverse, comoving separation vector of a neighbour light ray from
the fiducial ray, where $f_K(\chi)$ denotes the comoving angular
diameter distance and $\vec{\theta}$ the angular separation of the
light ray on the sky. We assume a flat sky and denote angular
positions by Cartesian vectors
\mbox{$\vec{\theta}=\theta_1+\i\theta_2$} in a complex notation with
origin $\vec{\theta}=0$ in the direction of the fiducial ray. For
sources distributed along radial distance $\chi$ according to the
probability density function (PDF) $p_{\rm s}(\chi)$, the effective
convergence at $\vec{\theta}$ is a projection of
$\delta_{\rm m}(\vec{x})$ onto the sky:
\begin{align}
\label{eq:kappafunction}
\kappa (\vec{\theta}) &= \int_0^{\chi_{\rm h}}\d\chi\;
g(\chi)\,\deltam\Big(f_K(\chi)\,\vec{\theta},\chi\Big)\;;
\\
g(\chi)&= \frac{3H_0^2\,\Omega_{\rm
    m}}{2c^2}\,\frac{f_K(\chi)}{a(\chi)} \int_\chi^{\chi_{\rm
    h}}\d\chi^\prime\; p_{\rm s}(\chi^\prime)\,
\frac{f_K(\chi^\prime-\chi)}{f_K(\chi^\prime)}\;,
\label{eq:gweight}
\end{align}
where $c$ is the vacuum speed of light, and $H_0$ is the Hubble
constant; $a(\chi)$ is the scale factor with $a(\chi)=1$ for $\chi=0$;
and $\chi_{\rm h}$ is the radius of the observable Universe
\citep{schneider2006gravitational}.  The convergence is related to the
(Cartesian) shear field $\gamma_{\rm c}(\vec{\theta})$ up to a
constant $\kappa_0$ through the convolution integral
\begin{equation}
  \label{eq:ks93}
  \kappa(\vec{\theta})-\kappa_0=
  \frac{1}{\pi}\int\d^2\theta^\prime\;
  {\cal D}^\ast(\vec{\theta}-\vec{\theta}^\prime)\,
  \gamma_{\rm c}(\vec{\theta}^\prime)
  =: 
  \kappa_{\rm E}(\vec{\theta})+\i\kappa_{\rm B}(\vec{\theta})\;,
\end{equation}
with the kernel
\begin{equation}
  {\cal D}^\ast(\vec{\theta})=
  \frac{\theta_2^2-\theta_1^2+2\i\theta_1\theta_2}{|\vec{\theta}|^4}
\end{equation}
\citep{1993ApJ...404..441K}.  We call the real part
$\kappa_{\rm E}(\vec{\theta})$ of the convergence the E-mode of the
convergence field and the imaginary part
$\kappa_{\rm B}(\vec{\theta})$ its B-mode. In the weak-lensing regime,
lens-lens couplings are negligible \citep{2009A&A...499...31H} so that
B-modes serve as indicator of systematic errors for our lensing
analysis. In practical studies, we exploit that measurements of the
image ellipticity of source galaxies can be converted into unbiased
estimates of $\gamma_{\rm c}$ at the positions $\vec{\theta}$ of the
sources if \mbox{$|\kappa|\ll1$}
\citep[e.g.,][]{2016arXiv160907937S}. In the following, we therefore
denote by $\epsilon_i$ an unbiased estimator of
$\gamma_{\rm c}(\vec{\theta}_i)$ at the position $\theta_i$ of a
source galaxy.

We consider the tangential- and cross-shear components
$\gamma_{\rm t}$ and $\gamma_\times$ of $\gamma_{\rm c}(\vec{\theta})$
relative to an orientation angle $\varphi$, namely we define the
$\varphi$-rotated shear,
\begin{equation}
  \label{eq:gammarot}
  \gamma(\vec{\theta};\varphi)=
  -\e^{-2\i\varphi}\,\gamma_{\rm c}(\vec{\theta})\;,
\end{equation}
and its decomposition
\begin{equation}
  \gamma_{\rm t}(\vec{\theta};\varphi)+
  \i\gamma_\times(\vec{\theta};\varphi):=
  \gamma(\vec{\theta};\varphi)\;.
\end{equation}
Also, for mathematical convenience, we denote differences between two
vectors $\vec{\theta}_i$ and $\vec{\theta}_j$ on the flat sky by
\begin{equation}
  \vec{\theta}_{ij}:=
  \vec{\theta}_i-\vec{\theta}_j=
  \theta_{ij}\,\e^{\i\varphi_{ij}}
\end{equation}
with the polar coordinates $\theta_{ij}$ and $\varphi_{ij}$.

\subsection{Galaxy clustering and galaxy-galaxy lensing}

In galaxy-galaxy lensing, we correlate positions of lens galaxies with
the tangential shear around these galaxies. For this, consider the
number density $N_{\rm g}(\vec{\theta})$ of lens galaxies on the sky
and their density contrast
\begin{equation}
\label{eq:1}
\kappagal(\vec{\theta})
= \frac{N_{\rm g}(\vec{\theta})-\overline{N}_{\rm g}}{\overline{N}_{\rm g}}
\end{equation}
relative to the mean number density $\overline{N}_{\rm g}$. Similar to
$\kappa(\vec{\theta})$, the density contrast
$\kappa_{\rm g}(\vec{\theta})$ constitutes a projection along the
line-of-sight $\vec{\theta}$. Specifically, let
$\delta_{\rm g}(\vec{x})$ be the relative fluctuations
$\delta n_{\rm g}/\overline{n}_{\rm g}$ in the three dimensional
number density of galaxies at a position relative to the fiducial
light ray and $p_{\rm d}(\chi)$ the PDF of galaxy distances $\chi$
inside the observed light cone. Then the density contrast of lenses on
the sky is
\begin{equation}
  \kappagal(\vec{\theta}) = 
  \int_0^{\chi_{\rm h}}\d\chi\;p_{\rm d}(\chi)\,\deltagal\Big(f_K(\chi)\vec{\theta},\chi\Big)
\end{equation}
\citep[e.g.,][]{2002ApJ...577..604H}.

Following \citet{1980lssu.book.....P}, we quantify the second-order
angular clustering of lenses by the correlation of two density
contrasts with separation $\vartheta$,
\begin{equation}
  \omega(\vartheta)=
  \Ave{\kappa_{\rm g}(\vec{\theta})\kappa_{\rm g}(\vec{\theta}+\vec{\vartheta})}\;.
\end{equation}
Owing to statistical isotropy and homogeneity of the random field
$\delta_{\rm g}$, the correlation function $\omega(\vartheta)$ depends
only on the separation $\vartheta$ of the two points. Likewise, we
assume isotropy and homogeneity for the matter-density
fluctuations $\delta_{\rm m}$ so that all following functions that
correlate two or more points depend only on the mutual separation of
points.

For a cross-correlation of lens positions with the projected matter
density, we define the correlation between the number density of
lenses and the mean tangential shear 
\begin{equation}
  \label{eq:ggl}
  \overline{\gamma}_{\rm t}(\vartheta)=
  \frac{1}{\overline{N}_{\rm g}}
  \Ave{N_{\rm g}(\vec{\theta})\,\gamma(\vec{\theta}+\vec{\vartheta};\varphi)}
  =\Ave{\kappa_{\rm g}(\vec{\theta})\,\gamma(\vec{\theta}+\vec{\vartheta};\varphi)}
\end{equation}
with $\varphi$ being the polar angle of
$\vec{\vartheta}=\vartheta\,\e^{\i\varphi}$. The imaginary part or
cross-component $\overline{\gamma}_\times(\vartheta)$ of this
correlator vanishes in the statistical average because of a parity
invariance of the random fields
\citep{2003A&A...408..829S}. Physically, galaxy-galaxy lensing probes
the (axis-symmetric) profile of the stacked (i.e., ensemble-averaged)
convergence
$\overline{\kappa}(\vartheta)=\ave{\kappa_{\rm
    g}(\vec{\theta})\kappa(\vec{\theta}+\vec{\vartheta})}$
around lenses at separation $\vartheta$:
\begin{equation}
  \overline{\gamma}_{\rm t}(\vartheta)=
  \left(
    \frac{2}{\vartheta^2}
    \int_0^\vartheta\d\vartheta^\prime\;\vartheta^\prime\,\overline{\kappa}(\vartheta^\prime)
  \right)
  -\overline{\kappa}(\vartheta)
\end{equation}
\citep{1995ApJ...439L...1K}.  

\subsection{Galaxy-galaxy-galaxy lensing}

\begin{figure}
  \centerline{\includegraphics[width=80mm]{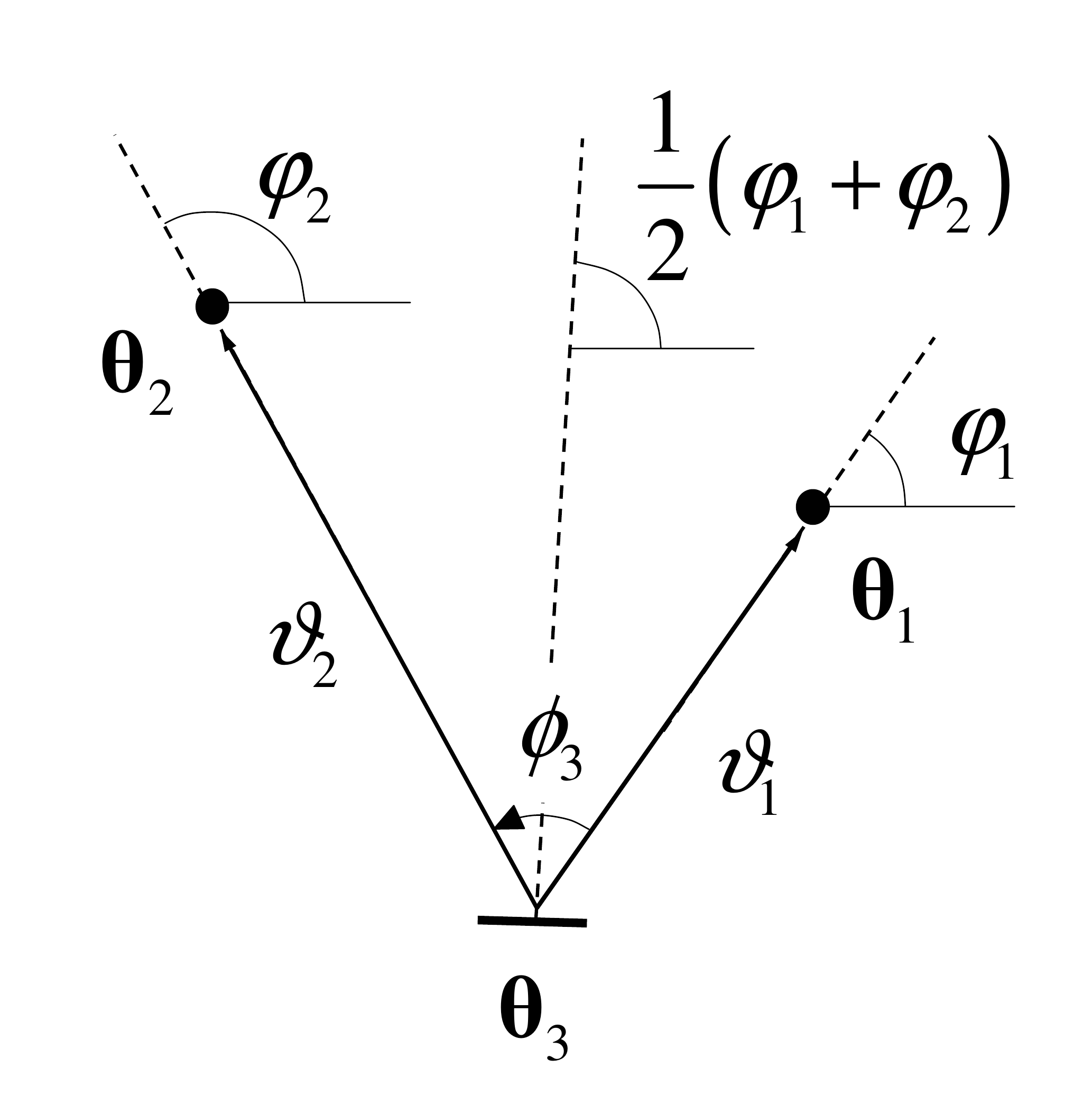}}
  \caption{\label{fig:g3lsketch} Geometry in the galaxy-galaxy-galaxy
    lensing correlation function $\cal G$ for lenses at
    $\vec{\theta}_1$ and $\vec{\theta}_2$, and a source at
    $\vec{\theta}_3$. The separations between source and lenses are
    $\vartheta_1$ and $\vartheta_2$. The separations of the lenses is
    $\theta_{12}=|\vec{\theta}_1-\vec{\theta}_2|$. This figure is
    copied from \cite{2005A&A...432..783S}.}
\end{figure}

A correlation function similar to galaxy-galaxy lensing can be defined
by measuring the average tangential shear at $\vec{\theta}_3$ around a
pair of lenses at $\vec{\theta}_1$ and $\vec{\theta}_2$. This is
introduced as one of two correlation functions of galaxy-galaxy-galaxy
lensing in \citet{2005A&A...432..783S} by
\begin{multline}
  \label{eq:gcorrelator}
  \mathcal{G}(\vartheta_1, \vartheta_2, \phi_3) =
  \Ave{\kappagal(\vtheta_1)\,\kappagal(\vtheta_2)\,
    \gamma\eb{\vtheta_3 ; \frac{\varphi_1+\varphi_2}{2}}}
  \\
  =-\e^{-\i(\varphi_1+\varphi_2)}\,
  \Ave{\kappagal(\vtheta_1)\,\kappagal(\vtheta_2)\, \gamma_{\rm
      c}(\vtheta_3)}\;,
\end{multline}
where $\phi_3=\varphi_2-\varphi_1$ denotes the angle spanned by the
two separation vectors $\vec{\theta}_1-\vec{\theta}_3$ and
$\vec{\theta}_2-\vec{\theta}_3$ with polar angles $\varphi_1$ and
$\varphi_2$ respectively. Figure \ref{fig:g3lsketch} sketches the
geometry. The tangential shear is thus defined relative to the line
that bisects the angle between the two lenses.

From a mathematical point of view, $\cal G$ is a correlator between
the random field $\kappa_{\rm g}(\vec{\theta})$ at $\vec{\theta}_1$
and $\vec{\theta}_2$, and the random field
$\gamma_{\rm c}(\vec{\theta})$ at $\vec{\theta}_3$. Since both fields
vanish on average, i.e.,
$\ave{\kappa_{\rm g}(\vec{\theta})}=\ave{\gamma_{\rm
    c}(\vec{\theta})}=0$
for all $\vec{\theta}$, the correlation function $\cal G$ contains no
additive contributions from correlators smaller than third order (the
so-called unconnected terms) and thus vanishes for purely Gaussian
random fields \citep[e.g.,][]{2010gfe..book.....M}.

\section{Mapping of mass correlated with lens pairs}
\label{sect:ExcessMass}

\subsection{Definitions of the excess mass}

In the case of galaxy-galaxy lensing, the correlator
$\overline{\gamma}_{\rm t}$ can be related to the stacked convergence
around individual lenses. This is modified in the case of
galaxy-galaxy-galaxy lensing, whereby the correlator $\cal G$ can now
be related to the stacked convergence around individual pairs of
lenses \citep{2008A&A...479..655S,2013MNRAS.430.2476S}. We show this
by considering first the shear pattern around an average lens pair.
Stacking the shear field around two lenses at $\vec{\theta}_1$ and
$\vec{\theta}_2$ results in the average
\begin{multline}
\label{eq:kkgamma}
\frac{\Ave{N_{\rm g}(\vtheta_1)N_{\rm g}(\vtheta_2)\gamma_{\rm
      c}(\vtheta_3)}} {\Ave{N_{\rm g}(\vtheta_1)N_{\rm g}(\vtheta_2)}}
= \frac{\Ave{[1+\kappa_{\rm g}(\vtheta_1)][1+\kappa_{\rm
      g}(\vtheta_2)]\gamma_{\rm c}(\vtheta_3)}}
{1+\omega(\theta_{12})}\\
= \frac{\Ave{\kappa_{\rm g}(\vtheta_1)\,\kappa_{\rm
      g}(\vtheta_2)\gamma_{\rm c}(\vtheta_3)}+ \Ave{\kappa_{\rm
      g}(\vtheta_1)\,\gamma_{\rm c}(\vtheta_3)}+\Ave{\kappa_{\rm
      g}(\vtheta_2)\, \gamma_{\rm
      c}(\vtheta_3)}}{1+\omega(\theta_{12})}\;,
\end{multline}
where we have used
$\ave{N_{\rm g}(\vec{\theta})N_{\rm
    g}(\vec{\theta}+\vec{\vartheta})}=\overline{N}_{\rm
  g}^2(1+\omega(\vartheta))$, 
and define $\theta_{12}$ as the separation of the lenses. In order to understand
why the left-hand side of \Ref{eq:kkgamma} equals the average shear
around two lens galaxies, we consider the number-density field
of lenses projected on a regular grid with a large number of
micro-cells, each with a solid angle $\sigma$. We define the micro-cells such 
that they are sufficiently fine as to contain at most one lens, such that 
$\int_\sigma\d\sigma\,N_{\rm g}(\vec{\theta})=0$ or 1. We then
have a contribution to
$\ave{N_{\rm g}(\vec{\theta}_1)N_{\rm g}(\vec{\theta}_2)\gamma_{\rm
    c}(\vec{\theta}_3)}$
only for
$\int_\sigma\d\sigma\,N_{\rm
  g}(\vec{\theta}_1)=\int_\sigma\d\sigma\,N_{\rm
  g}(\vec{\theta}_2)=1$,
while $\ave{N_{\rm g}(\vec{\theta}_1)N_{\rm g}(\vec{\theta}_2)}$ in
the denominator is the probability to have a pair of galaxies at
$\vec{\theta}_1$ and $\vec{\theta}_2$ at the same time (and is therefore a normalisation
factor).

Now, through the definitions \Ref{eq:gammarot} and \Ref{eq:ggl} for
galaxy-galaxy lensing we additionally have
$\ave{\kappa_{\rm g}(\vec{\theta}_j)\gamma_{\rm
    c}(\vec{\theta}_3)}=-\e^{2\i\varphi_{j3}}\,\overline{\gamma}_{\rm
  t}(\vartheta_j)$ and can therefore cast \Ref{eq:kkgamma} into
\begin{eqnarray}
  \nonumber
  \Ave{\kappa_{\rm g}(\vtheta_1)\kappa_{\rm g}(\vtheta_2)\gamma_{\rm
      c}(\vtheta_3)}&=&
  \Big(1+\omega(\theta_{12})\Big)\, \frac{\ave{N_{\rm
        g}(\vtheta_1)N_{\rm g}(\vtheta_2)\gamma_{\rm c}(\vtheta_3)}}
  {\ave{N_{\rm g}(\vtheta_1)N_{\rm g}(\vtheta_2)}}
  \\
  \label{eq:kkgamma1}
  &+&\e^{2\i\varphi_{13}}\,\overline{\gamma}_{\rm t}(\vartheta_1)
  +\e^{2\i\varphi_{23}}\,\overline{\gamma}_{\rm t}(\vartheta_2)\;.
\end{eqnarray}
This shows that $\cal G$ is, apart from a phase factor, indeed related
to the average shear around lens pairs given by the first term on the
right-hand side -- but rescaled with $1+\omega(\theta_{12})$ and in
excess of the mean shear around individual lenses as given by the two
terms that involve $\overline{\gamma}_{\rm t}$.

\begin{figure}
  \centerline{\includegraphics[width=85mm,trim=50mm 120mm 30mm 30mm,clip=true]{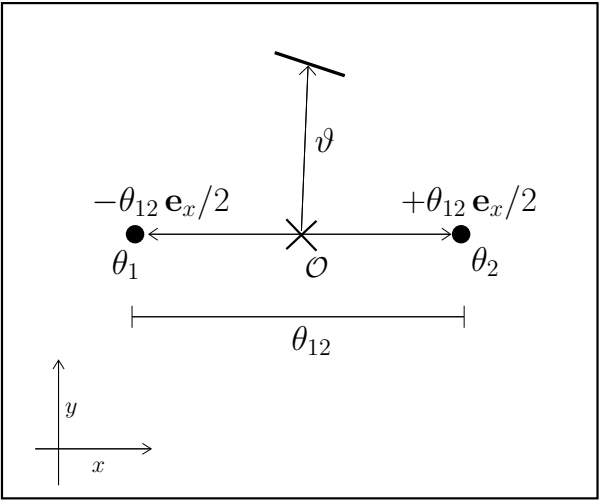}}
  \caption{\label{fig:stackgeometry} Cartesian coordinate frame for a
    map of excess mass or the pair convergence. Lenses with separation
    $\theta_{12}$ are at the positions as indicated by the solid
    points. A location inside the map is identified by the
    two-dimensional vector $\vec{\vartheta}$.}
\end{figure}

For the excess mass map, we consider a two-dimensional convergence map
$\overline{\Delta\kappa}_{\rm emm}(\vec{\vartheta};\theta_{12})$ that
corresponds to the excess shear in Eq. \Ref{eq:kkgamma1} around lenses
at given separation $\theta_{12}$. We construct this map in a specific
coordinate frame, for which $\vec{\vartheta}$ is the relative
separation from the map centre $\cal O$.  See the sketch in
Fig. \ref{fig:stackgeometry}.  The lenses are located at
$\vec{\theta}_1=-\theta_{12}\,\vec{e}_x/2$ and
$\vec{\theta}_2=+\theta_{12}\,\vec{e}_x/2$ with $\vec{e}_x$ being a
unit vector in the $x$-direction. Applying the linear Kaiser-Squires
transformation in Eq. \Ref{eq:ks93} to Eq. \Ref{eq:kkgamma1} in this
coordinate frame yields, up to a constant $\kappa_0$, a convergence
map that can be expanded according to
\begin{multline}
  \label{eq:kkkappa}
  \overline{\Delta\kappa}_{\rm emm}(\vec{\vartheta};\theta_{12})
  \\
  =\Big(1+\omega(\theta_{12})\Big)\,
  \overline{\kappa}_{\rm pair}(\vec{\vartheta};\theta_{12})
  -\overline{\kappa}_{\rm ind}(|\vec{\vartheta}-\vec{\theta}_1|)
  -\overline{\kappa}_{\rm ind}(|\vec{\vartheta}-\vec{\theta}_2|)+\kappa_0\;.
\end{multline}
Here the average convergence
$\overline{\kappa}_{\rm pair}(\vec{\vartheta};\theta_{12})$ around
lens pairs corresponds to the shear stack
$\ave{N_{\rm g}N_{\rm g}\gamma_{\rm c}}/\ave{N_{\rm g}N_{\rm g}}$ in
\Ref{eq:kkgamma1}, and the average convergence
\mbox{$\overline{\kappa}_{\rm ind}(|\vec{\vartheta}|)=\ave{N_{\rm
      g}(\vec{\theta})\kappa(\vec{\theta}+\vec{\vartheta})}/\overline{N}_{\rm
    g}$}
around individual lenses corresponds to the average shear
$-\e^{2\i\varphi}\,\overline{\gamma}_{\rm t}(|\vec{\vartheta}|)$,
centred on the location of each lens. We emphasise that
$1+\omega(\theta_{12})$ is a constant in this map, and that
$\overline{\kappa}_{\rm ind}(|\vec{\vartheta}|)$ is, by definition of
$\cal G$, the average convergence around \ita{all} galaxies in the
sample -- including those that do not have a partner at separation
$\theta_{12}$.

Since the relation between $\kappa(\vec{\theta})$ and
$\gamma_{\rm c}(\vec{\theta})$ is only defined up to a constant
$\kappa_0$, we cannot uniquely determine the excess mass map from the
excess shear (see Eq. \ref{eq:ks93}). It is, however, reasonable to
assume that
$\overline{\Delta\kappa}_{\rm emm}(\vec{\vartheta};\theta_{12})$,
being the three-point correlation function
$\ave{\kappa_{\rm g}\kappa_{\rm g}\kappa}$, quickly approaches zero
for large $\vartheta$ which might be used to define
$\kappa_0$. Alternatively, for the maps presented here, we fix
$\kappa_0$ by asserting that $\overline{\Delta\kappa}_{\rm emm}$
vanishes when averaged over the entire map area. We will neglect
$\kappa_0$ in the following equations for convenience.

For what follows we also consider the pair convergence of lens 
pairs as a variant of the excess mass map, which is the straightforward
difference signal
\begin{multline}
  \label{eq:pc}
  \overline{\Delta\kappa}(\vec{\vartheta};\theta_{12}):=
  \overline{\kappa}_{\rm pair}(\vec{\vartheta};\theta_{12})-
  \overline{\kappa}_{\rm ind}(|\vec{\vartheta}-\vec{\theta}_1|)-
  \overline{\kappa}_{\rm ind}(|\vec{\vartheta}-\vec{\theta}_2|)
  \\
  =\overline{\Delta\kappa}_{\rm emm}(\vec{\vartheta};\theta_{12})
  -\omega(\theta_{12})\,\overline{\kappa}_{\rm
    pair}(\vec{\vartheta};\theta_{12})
\end{multline}
between the stacked convergence around lens pairs and the stacked
convergence around two individual lenses. Since the excess mass map,
originating from the connected correlation function $\cal G$, is free
of unconnected correlations (by definition), our interpretation is that
the excess mass is the connected part of the pair convergence, and the
extra term
$-\omega(\theta_{12})\,\overline{\kappa}_{\rm
  pair}(\vec{\vartheta};\theta_{12})$
is the unconnected part of the pair convergence.

While the excess mass exactly vanishes for Gaussian random fields, the
pair convergence generally does not; although it is entirely determined
by second-order correlations in this case. This can be seen from the
definition \Ref{eq:pc} of $\overline{\Delta\kappa}$ and
Eq. \Ref{eq:kkkappa} with
\mbox{$\overline{\Delta\kappa}_{\rm emm}=0$}, giving
\begin{equation}
  \overline{\Delta\kappa}(\vec{\vartheta};\theta_{12})
  =-\frac{\omega(\theta_{12})}
  {1+\omega(\theta_{12})}
  \Big(
  \overline{\kappa}_{{\rm
      ind}}(|\vec{\vartheta}-\vec{\theta}_1|)+
  \overline{\kappa}_{{\rm ind}}(|\vec{\vartheta}-\vec{\theta}_2|)
  \Big)\;.  
\end{equation}

We visualise the excess mass as a two-dimensional map by plotting
either $\overline{\Delta\kappa}_{\rm emm}$, or
$\overline{\Delta\kappa}$ for the pair convergence, as function of
$\vec{\vartheta}$ for a fixed lens-lens separation $\theta_{12}$ and
orientation. The resulting maps have two known symmetries
\citep{2008A&A...479..655S}. First, there is a parity symmetry:
correlation functions are unchanged under a reflection of shear and
the lens density across an axis owing to the parity invariance of
cosmological fields \citep{2003A&A...408..829S}. As a consequence,
quadrants in the maps are statistically consistent when mirrored
across the line connecting two lenses. Second, another symmetry is
present because we correlate density fluctuations $\kappa_{\rm g}$ at
$\vec{\theta}_1$ and $\vec{\theta}_2$ from the same galaxy sample: a
permutation of lens indices results in the same correlation
function. These symmetries combine to produce, in the absence of noise, an exact
reflection symmetry of the map with respect to both the $x$- and
$y$-axes. We exploit this symmetry to enhance the S/N in the maps by
averaging the quadrants inside each map.

Physically, the dimensionless quantities
$\overline{\Delta\kappa}_{\rm emm}$ and $\overline{\Delta\kappa}$ are
surface-mass densities in units of the (average) critical density
$\overline{\Sigma}_{\rm crit}:=1/\overline{\Sigma_{\rm crit}^{-1}}$,
defined by
\begin{multline}
  \label{eq:sigmacrit}
  \overline{\Sigma_{\rm crit}^{-1}}:=
  \\
  \frac{4\pi G_{\rm N}}{c^2} \int_0^{\chi_{\rm h}}\d\chi_{\rm
    s}\,\d\chi_{\rm d}\, p_{\rm d}(\chi_{\rm d})\,p_{\rm s}(\chi_{\rm
    s})\, \frac{a(\chi_{\rm d})f_K(\chi_{\rm d})f_K(\chi_{\rm
      s}-\chi_{\rm d})}{f_K(\chi_{\rm s})}\;,
\end{multline}
where $G_{\rm N}$ is Newton's gravitational constant. In our analysis,
lenses have a typical distance of \mbox{$z_{\rm d}\approx0.4$} and
sources of \mbox{$z_{\rm s}\approx0.93$} so that we estimate
\mbox{$\overline{\Sigma}_{\rm
    crit}\approx4.25\times10^3\,h\,\msol\,{\rm pc}^{-2}$}
as the fiducial value for the critical surface mass density in our
analysis below.

\subsection{Impact of chance galaxy pairs}
\label{sect:chancepairs}

The following is a simple discussion exploring the impact of
uncorrelated lens pairs on excess mass or pair convergence maps. In
the construction of our maps, we select galaxy pairs within a sample
by their angular separation $\theta_{12}$ on the sky. Therefore, there
will be pairs that are well separated in radial distance from each
other, such that third-order correlations involving these are
negligible for practical purposes. On the other hand, we will have
pairs which have non-negligible third-order correlations with the
lensing convergence as they are at similar distance. Naturally,
however, making a distinction between what is negligible or otherwise
is somewhat arbitrary. Nonetheless we could define a reasonable
threshold for the correlation amplitude or a maximum radial separation
of galaxies in a pair and use this to restrict our sample to those
pairs with higher expected S/N.  For the purpose of this simple
discussion, however, we assume that we have a sample of lenses with a
fraction $1-p_{\rm tp}$ of pairs for which a stack of convergence
shall be free of any third-order correlations; thus giving the
expectation value
\begin{equation}
 \label{eq:cp}
 \overline{\kappa}_{\rm
   pair}(\vec{\vartheta};\theta_{12})=
 \overline{\kappa}_{\rm ind}(|\vec{\vartheta}-\vec{\theta}_1|)+
 \overline{\kappa}_{\rm ind}(|\vec{\vartheta}-\vec{\theta}_2|)=:
 \overline{\kappa}_{\rm
   pair}(\vec{\vartheta};\theta_{12})|_{\rm cp}\;.
\end{equation}
These are the `chance pairs'; sources which are pairs only in
projection on-sky.  The remaining fraction $p_{\rm tp}$ of (physically
connected) pairs, to which we refer as `true pairs', shall have a
different yet unspecified stack
\mbox{$\overline{\kappa}_{\rm
    pair}(\vec{\vartheta};\theta_{12})=\overline{\kappa}_{\rm
    pair}(\vec{\vartheta};\theta_{12})|_{\rm tp}$}
that depends in detail on the average surface mass-density around
those pairs and the critical density $\overline{\Sigma}_{\rm crit}$ at
the distance of the pair. The stack around true pairs carries the
interesting physical information so that, ideally, we would like to
define an excess mass that is independent from chance pairs. This is
neither true for an excess mass map nor for a pair convergence map.

We start by exploring the behaviour of pure samples of chance or true
pairs.  For pure samples of chance pairs (i.e., \mbox{$p_{\rm tp}=0$})
the excess mass and pair convergence vanish because mass cannot be
correlated with two statistically independent lenses.  Indeed, we will
find
\mbox{$\overline{\kappa}_{\rm
    pair}(\vec{\vartheta};\theta_{12})=\overline{\kappa}_{\rm
    pair}(\vec{\vartheta};\theta_{12})|_{\rm cp}$},
\mbox{$\omega(\theta_{12})=0$}, and consequently
\mbox{$\overline{\Delta\kappa}_{\rm emm}=\overline{\Delta\kappa}=0$}.
On the other hand, if we have a pure sample of true pairs
($p_{\rm tp}=1$) with a clustering amplitude $\omega_{\rm tp}$ (at
separation $\theta_{12}$), we will find
\begin{eqnarray}
  \nonumber
  \lefteqn{\overline{\Delta\kappa}_{\rm emm}(\vec{\vartheta};\theta_{12})=}
  \\
  \nonumber
  &&\!\!\!\!\Big(1+\omega_{\rm tp}\Big)\, \overline{\kappa}_{\rm
    pair}(\vec{\vartheta};\theta_{12})|_{\rm
    tp}-\overline{\kappa}_{\rm ind}(|\vec{\vartheta}-\vec{\theta}_1|)-
  \overline{\kappa}_{\rm ind}(|\vec{\vartheta}-\vec{\theta}_2|)
  \\
   \label{eq:tpemm}
  &&\!\!\!\!=:\overline{\Delta\kappa}_{\rm emm}(\vec{\vartheta};\theta_{12})|_{\rm
    tp}
\end{eqnarray}
for the excess mass. We obtain the equation for
$\overline{\Delta\kappa}(\vec{\vartheta};\theta_{12})$ by setting
$\omega_{\rm tp}$ to zero in \Ref{eq:tpemm}.

Usually we have a mixture of chance pairs and true pairs, and the
impact of chance pairs on the excess mass is not immediately
obvious. To explore this case, let us now assume that
\mbox{$0<p_{\rm tp}<1$} and that the mean tangential shear
$\overline{\gamma}_{\rm t}(\vartheta)$ is as unchanged from the pure case. 
In this mixture, the clustering amplitude is reduced to
\mbox{$\omega=p_{\rm tp}\,\omega_{\rm tp}$}, and the convergence stack
around all pairs,
\begin{equation}
  \label{eq:stacksplit}
  \overline{\kappa}_{\rm pair}(\vec{\vartheta};\theta_{12})=
  p_{\rm tp}\,\overline{\kappa}_{\rm pair}(\vec{\vartheta};\theta_{12})|_{\rm tp}+
  (1-p_{\rm tp})\,\overline{\kappa}_{\rm pair}(\vec{\vartheta};\theta_{12})|_{\rm cp}\;,
\end{equation}
is the weighted average of the stacks in the pure samples. Then using
\Ref{eq:cp}-\Ref{eq:stacksplit} in Eq. \Ref{eq:kkkappa} results, after
some algebra, in the excess mass of a mixed sample:
\begin{eqnarray}
  \nonumber
  \lefteqn{\overline{\Delta\kappa}_{\rm emm}(\vec{\vartheta};\theta_{12})=}
  \\
  \nonumber
  &&p_{\rm tp}\,\frac{1+p_{\rm tp}\,\omega_{\rm tp}}
  {1+\omega_{\rm tp}}\;
  \overline{\Delta\kappa}_{\rm emm}(\vec{\vartheta};\theta_{12})|_{\rm tp}
  \\
  \label{eq:mix1b}
  &&+p_{\rm tp}\,(1-p_{\rm
    tp})\,\frac{\omega_{\rm tp}^2}{1+\omega_{\rm tp}}\,\Big(
  \overline{\kappa}_{\rm ind}(|\vec{\vartheta}-\vec{\theta}_1|)+
  \overline{\kappa}_{\rm ind}(|\vec{\vartheta}-\vec{\theta}_2|)
  \Big)\;.
\end{eqnarray}
In conclusion, while the presence of chance pairs merely diminishes
the overall amplitude of the true pair excess-mass
$\overline{\Delta\kappa}_{\rm emm}(\vec{\vartheta};\theta_{12})|_{\rm
  tp}$
in the mixture, there is also a second term in the last line of
Eq. \Ref{eq:mix1b} that changes the overall appearance of the map by
adding an extra signal, mainly at the lens positions, that is
proportional to the convergence stack around chance pairs.

This extra signal can be avoided in the pair convergence. Namely, by
plugging \Ref{eq:cp} and \Ref{eq:stacksplit} into Eq. \Ref{eq:pc} we
get
\begin{equation}
  \label{eq:paircon2}
  \overline{\Delta\kappa}(\vec{\vartheta};\theta_{12})=
  p_{\rm tp}\,\Big( \overline{\kappa}_{\rm
    pair}(\vec{\vartheta};\theta_{12})|_{\rm tp}- \overline{\kappa}_{\rm
    ind}(|\vec{\vartheta}-\vec{\theta}_1|)- \overline{\kappa}_{\rm
    ind}(|\vec{\vartheta}-\vec{\theta}_2|) \Big)
\end{equation}
for a mixture sample. The presence of chance pairs does at most change
the overall amplitude in the pair convergence; each value in the map
gives a lower limit to the pair convergence around true pairs inside
the brackets of \Ref{eq:paircon2}.

\section{Data}
\label{sect:data}

\subsection{The Canada-France-Hawaii Telescope Lensing Survey}
\label{sect:CFHTLenSdata}

The CFHTLenS is a multi-colour, wide-field lensing survey with
measurements of galaxy photometry in the five bands
$u^\ast g^\prime r^\prime i^\prime z^\prime$ (AB system), observed as
part of the CFHT Legacy Survey Wide \citep{2012MNRAS.427..146H}.  The
survey covers $154\,\mathrm{deg}^2$ of the sky, consisting of four
contiguous fields: W1 ($\approx64\,\rm deg^2$), W2
($\approx23\,\rm deg^2$), W3 ($\approx44\,\rm deg^2$), and W4
($\approx23\,\rm deg^2$). The seeing is optimised for lensing
measurements in the $r^\prime$-band and lies typically between
$0\overset{\pprime}{.}66-0\overset{\pprime}{.}82$; the camera
resolution is $0\overset{\pprime}{.}187$ per CCD pixel. Each field is
a mosaic of a set of Megacam pointings with $1\times1\,\rm deg^2$
field-of-view each. The data reduction uses the processing pipeline
\texttt{THELI} \citep{2013MNRAS.433.2545E}, shear measurements of
source galaxies are made using \ita{lens}fit
\citep{2013MNRAS.429.2858M}, and estimates for galaxy photometric
redshifts and stellar masses are obtained using PSF-matched photometry
and the computer code \texttt{BPZ} (\citealt{2012MNRAS.421.2355H},
\citealt{2000ApJ...536..571B}). The photometric estimator $z_{\rm ph}$
for the redshift is the maximum in the posterior redshift distribution
of a galaxy. The final galaxy catalogue comprises these physical
parameters of $7\times10^6$
objects.\footnote{\url{http://www.cadc-ccda.hia-iha.nrc-cnrc.gc.ca/en/community/CFHTLens/query.html}}
For our analysis, we use only Megacam pointings that are flagged as
`good' for lensing studies. This amounts to roughly 75 per cent of the
pointings with an effective area of $A\approx95\,\rm deg^2$, which we
determined by counting the number of unmasked pixels in the mask files
of the 129 good pointings. For the selection of lens and source
catalogues we follow the criteria in S13, where additionally several
tests for systematic errors in the galaxy-galaxy-galaxy lensing
measurements (for our samples) have already been performed. S13 also
shows plots of the redshift distributions of galaxy samples.

Our source samples are galaxies with \mbox{$i^\prime<24.7$},
photometric redshifts of \mbox{$0.65\le z_{\rm ph}<1.2$}, and
non-vanishing statistical weights $w_i$ according to
\ita{lens}fit. This gives roughly $2.2\times10^6$ sources inside the
good pointings, and a mean $w$-weighted redshift of
\mbox{$\bar{z}_{\rm s}\approx0.93$}. The lower limit in $z_{\rm ph}$
is chosen to reduce the overlap in redshift between lens and source
samples in order to suppress systematic errors in the correlation
functions. The effective number density of our sources is
$\overline{N}_{\rm s}=(\sum_iw_i)^2/(A\,\sum_iw_i^2)\approx5.5\,\rm
arcmin^{-2}$.
To correct for the additive and multiplicative bias $m_i$ in the
estimators $\epsilon_i$ of the shear of the $i$th source, we follow
the instructions in \citet{2013MNRAS.429.2858M}. For this, the
correction for $m_i$ is easily included in our shear-related
estimators by replacing $\epsilon_i\mapsto\epsilon_i\,(1+m_i)^{-1}$
and $w_i\mapsto w_i\,(1+m_i)$. Note that we always have $m_i>-1$ (see
Appendix A in S13).

For the lens samples, we select galaxies with flux limit
$i^\prime\le22.5$ from two photo-$z$ bins:
\mbox{$0.2\le z_{\rm ph}<0.44$} (`low-$z$') and
\mbox{$0.44\le z_{\rm ph}<0.6$} (`high-$z$'). In addition, we select
galaxies only from the stellar-mass range between
\mbox{$5\times10^9\le M_{\rm sm}/{\rm M}_\odot<3.2\times10^{11}$}
which combines all stellar-mass samples sm1 to sm6 in S13. This
selection picks galaxies around the characteristic mass
\mbox{$M_\ast\approx5\times10^{10}\,\msol$} of the stellar-mass
function \citep[][and references therein]{2017MNRAS.470..283W}.  The
estimates for stellar masses assume an initial-mass function according
to \citet{2003PASP..115..763C} and have a typical RMS error of 0.3 dex
\citep{2014MNRAS.437.2111V}. Counting only good pointings, we have in
total $1.8\times10^5$ galaxies in the low-$z$ sample, yielding
\mbox{$\overline{N}_{\rm g}\approx0.5\,\rm arcmin^{-2}$}, and
$2.5\times10^5$ galaxies in the high-$z$ sample,
\mbox{$\overline{N}_{\rm g}\approx0.7\,\rm arcmin^{-2}$}. The mean
redshifts are $\bar{z}_{\rm d}\approx0.35,0.51$ for low-$z$ and
high-$z$ respectively. The RMS error of the redshift estimates is
\mbox{$\sigma(z)\approx0.04\,(1+z)$} with an outlier rate of roughly
$3\%$.

The lower limit of allowed stellar masses in addition to the
$i^\prime\le22.5$ flux limit makes our lens sample approximately
volume-limited inside the redshift intervals, as can be seen from
Fig. \ref{fig:completeness}. The figure shows absolute rest-frame
magnitudes and colours versus redshift for the CFHTLenS galaxies: the
blue dots are galaxies with stellar-mass selection, and the orange
dots are galaxies without stellar-mass selection. Only the few
galaxies that are fainter than \mbox{$M_u\gtrsim-18.5$} or
\mbox{$M_g\gtrsim-19.5$} are missing for
\mbox{$z_{\rm ph}\gtrsim0.4$}. By comparing the distribution of blue
and orange points in the colour plots we also see that the
stellar-mass selection rejects galaxies bluer than
\mbox{$M_g-M_r\lesssim0.3$} or \mbox{$M_u-M_g\lesssim0.6$} at all
redshifts (left column in figure).

\subsection{Synthetic lensing data and mock galaxies}
\label{sect:MockData}

Our mock lensing-data are generated by tracing the distortion of light
bundles that traverse 64 independent light cones of the $N$-body
Millennium Simulation (\citealt{2005Natur.435..629S};
\citealt{2009A&A...499...31H}). The Millennium Simulation is a purely
dark-matter simulation with a spatial (comoving) resolution of
$5\,h^{-1}\,\rm kpc$ that is sampled by $\sim10^{10}$ mass particles,
populating a cubic region of comoving side length of
$500\,h^{-1}\,\rm Mpc$. The fiducial cosmology of the Millenium
Simulation has the following parameters:
$\Omega_{\rm m}=0.25=1-\Omega_\Lambda$ for density parameters of
matter and dark energy; $\Omega_{\rm b}=0.045$ for the baryon density;
$\sigma_8=0.9$ for the normalisation of the linear matter power
spectrum at $z=0$; a Hubble parameter of
$H_0=h\,100\,\rm km\,s^{-1}\,Mpc^{-1}$ with $h=0.73$; and primordial
fluctuations as in the Harrison-Zeldovich model.  The fiducial
cosmology in the Millennium Simulation is somewhat different compared
to recent results by the \cite{2016A&A...594A..13P}. However, we
expect that this recent update in cosmological parameters only mildly
affects our model predictions, similar to the clustering properties of
SAM galaxies in \cite{2017MNRAS.469.2626H} where the authors rescale
the Millennium Simulation results to the \ita{Planck} cosmology for a
comparison.

Regarding the synthetic lensing data, the simulated light-bundle
distortions are an average over source distances with a probability
distribution that is the same as that of CFHTLenS sources (see Fig. 5
in S13). Each of the 64 simulated light cones yields a
$4^\circ\times4^\circ$ square with information on the theoretical
lensing convergence and shear along the line-of-sight at $4096^2$
pixel positions. The total area of the mock data is therefore
$1024\,\deg^2$. We mainly use the convergence grid for the estimator
in Sect. \ref{sect:EstimatorSAM}. Only where we compare the
convergence-stack method to the shear-stack method
(Sect. \ref{sect:EstimatorData}), or where we quantify the model
residuals in the maps, do we also generate synthetic shear
catalogues. For these catalogues, we uniformly pick random source
positions on the grid and assign each the shear value closest to that
position as the observed source ellipticity. We therefore do not
incorporate any shape noise.

For mock lens galaxies, we apply the H15 SAM-description, adjusted to
the fiducial cosmology of the Millennium Simulation. We then follow
the steps in S17 (specifically those listed in their Sect. 3.2), to
obtain the mock lens samples for low-$z$ and high-$z$, with a
selection function that is consistent with that of galaxies in
CFHTLenS, and including an emulation of statistical errors in both
stellar masses and photometric redshifts. In particular, the mock
samples have redshift distributions that are consistent with the
CFHTLenS samples.

\section{Estimators of excess mass}
\label{sect:methods}

We present two estimators for both, the excess mass map
(Eq. \ref{eq:kkkappa}) and the pair convergence map (Eq. \ref{eq:pc}),
which aim for different applications. In one version, we stack the
shear field around pairs of galaxies and carry out a convergence
reconstruction of the stack afterwards. This approach is suitable for
observational weak-lensing data where only estimates of (reduced)
shear are available for a set of discrete positions of source
galaxies. In another version, we stack the convergence on contiguous
grids directly, either for quick model predictions of the excess mass
or to assess the accuracy of the shear-stacking method.

For practical implementations of both estimators, we note that
frequently occurring phase factors $\e^{2\i\varphi_{ij}}$ of
separation vectors in our usual complex notation
$\vec{\theta}_{ij}=\theta_{ij}\,\e^{\i\varphi_{ij}}$ are easily
computed by
$\e^{2\i\varphi_{ij}}=\vec{\theta}_{ij}^{}/\vec{\theta}_{ij}^\ast$.

\subsection{Shear stack}
\label{sect:EstimatorData}

Let $\vtheta^{\rm d}_i$ be positions of $n_{\rm d}$ lens galaxies on
the sky, and $(\epsilon_i,w_i,\vtheta^{\rm s}_i)$ the details of
$n_{\rm s}$ source galaxies with ellipticities $\epsilon_i$,
statistical weights $w_i$, and positions $\vtheta^{\rm s}_i$. From
this, we compute the excess shear in Eq. \Ref{eq:kkgamma1} that is
based on both the shear stack around galaxy pairs -- the first term of
the right-hand side of this equation -- and the average shear around
individual galaxies, which are the other terms on the right-hand
side. Importantly, the excess shear is not identical to the average
shear around galaxy pairs; it is usually only a small fraction of the
latter.

The overall strategy for an estimator of the excess mass map is: (i)
we estimate the excess shear by stacking source ellipticities around
lens pairs in an appropriate reference frame and with weights
$1+\omega(\theta_{12})$ (first term in Eq. \ref{eq:kkgamma1}) and (ii)
add the terms that involve $\overline{\gamma}_{\rm t}(\vartheta)$ to
the stack (other terms); (iii) we apply the Kaiser-Squires inversion,
Eq.  \Ref{eq:ks93}, to obtain the excess mass map in
Eq. \Ref{eq:kkkappa}; (iv) finally, we subtract a constant $\kappa_0$
from the map. The computation of the pair convergence is only slightly
different as we explicitly set $\omega(\theta_{12})\equiv0$ in this
procedure. The following describes the details of the stacking and the
convergence reconstruction. Therein we assume that estimates of
$\omega(\vartheta)$ and $\overline{\gamma}_{\rm t}(\vartheta)$ are
already available; see the following section for estimators of those.

\begin{figure}
  \centerline{\includegraphics[width=80mm]{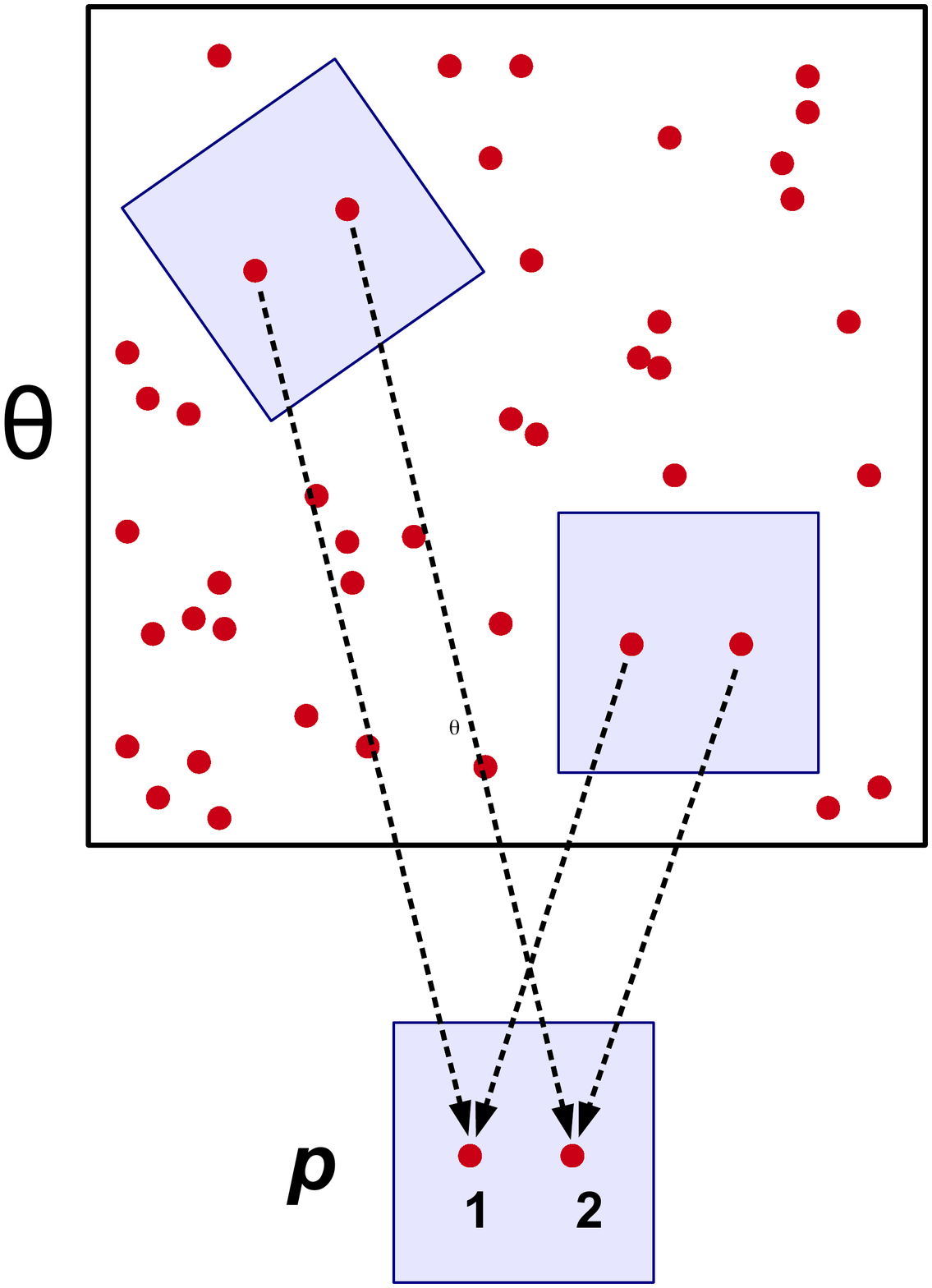}}
  \caption{\label{fig:xsmap} Schematic of the stacking
    procedure. Positions $\vec{\theta}$ on the sky are mapped to
    stacking frame positions $\vec{p}$ defined by the fixed positions
    of a selected lens pair (red points) in the stacking frame. The
    stacking-frame positions of the lenses are denoted by
    $\vec{p}^{\rm d}_1$ and $\vec{p}^{\rm d}_2$. The mapping involves
    a rotation so that source ellipticities in the stack have to be
    rotated.}
\end{figure}

For stacking, we define a two-dimensional grid with $N_{\rm p}\times
N_{\rm p}$ grid pixels (we choose $N_{\rm p}=200$); pixels shall have
a square geometry. Each grid cell $mn$ has the vector position
$\vec{p}_{mn}=m+{\rm i}\,n$ where $(m,n)$ are its coordinates in a
Cartesian stacking frame.  Let $(\vtheta_i^{\rm d},\vtheta_j^{\rm d})$
be the positions of a selected lens pair $ij$ within the separation
bin \mbox{$\vartheta-\Delta\vartheta/2\le|\vtheta_i^{\rm
    d}-\vtheta_j^{\rm d}|<\vartheta+\Delta\vartheta/2$}, and
$(\epsilon_k,w_k,\vtheta^{\rm s}_k)$ are the details of a source close
to the lens pair. We map $\vec{\theta}$-coordinates to
$\vec{p}$-coordinates by a rotation $\alpha_{ij}$ and scaling
$|A_{ij}|$, both encapsulated inside \mbox{$A_{ij}=|A_{ij}|\,\ee^{{\rm
      i}\alpha_{ij}}$}, and a translation $B_{ij}$,
\begin{equation}
  \vec{p}=A_{ij}\,\vec{\theta}+B_{ij}\;.
\end{equation}
The complex-valued parameters $A_{ij}$ and $B_{ij}$ are determined by
the mapping of the two lens positions to the fixed positions
$\vec{p}_1^{\rm d}$ and $\vec{p}_2^{\rm d}$ in the stack,
\begin{equation}
  \label{eq:affine}
  A_{ij}=\frac{\vec{p}_2^{\rm d}-\vec{p}_1^{\rm d}}
  {\vtheta_j^{\rm d}-\vtheta_i^{\rm d}}~;~
  B_{ij}=\frac{\vtheta_j^{\rm d}\,\vec{p}_1^{\rm d}-\vtheta_i^{\rm d}\,\vec{p}_2^{\rm d}}{\vtheta_j^{\rm d}-\vtheta_i^{\rm d}}\;
\end{equation}
(see Fig. \ref{fig:xsmap} for an illustration). The positions of
sources $\vtheta_k^{\rm s}$ are therefore
\mbox{$\vec{p}^{\rm s}_{ijk}=A_{ij}\,\vtheta^{\rm s}_k+B_{ij}$} in the
stacking frame. Additionally, the source ellipticities $\epsilon_k$
have to be rotated by
\mbox{$\epsilon_k\mapsto\e^{2\i\alpha_{ij}}\times\epsilon_k=A_{ij}^{}/A^\ast_{ij}\times\epsilon_k$}
in the $\vec{p}$-frame. Before mapping $\epsilon_k$ to the stacking
frame we subtract off the average shear around each lens position in
the $\vec{\theta}$-frame to obtain the excess shear. The complete
stack for the excess shear at the grid pixel $mn$ is then the weighted
sum
\begin{multline}
  \label{eq:excessshear}
\Delta\gamma_{mn}= 
\\
 \sum\limits_{i,j,k=1}^{n_{\rm d},n_{\rm s}}\!\!
   \frac{\Delta_{ijk}^{mn}\ w_k\,\ee^{2{\rm i}\alpha_{ij}}}{W_{mn}}\,
   \Big([1+\omega(\theta_{ij}^{\rm dd})]\,\epsilon_k
   +\ee^{2{\rm i}\varphi_{ik}}\,\overline{\gamma}_{\rm t}(\theta_{ik}^{\rm ds})
   +\ee^{2{\rm i}\varphi_{jk}}\,\overline{\gamma}_{\rm t}(\theta_{jk}^{\rm ds})\Big)\;,
\end{multline}
where the total weight is
\begin{equation}
  W_{mn}=\sum\limits_{i,j,k=1}^{n_{\rm d},n_{\rm s}}\Delta_{ijk}^{mn}\,w_k\;,
\end{equation}
and $\Delta_{ijk}^{mn}=1$ flags if the source position
$\vec{p}^{\rm s}_{ijk}$ falls within the grid cell $mn$ and
$\Delta_{ijk}^{mn}=0$ otherwise; by $\theta_{ij}^{\rm dd}$ we denote
the separation between the lenses $i$ and $j$, and by
$\vtheta_{ik}^{\rm ds}$ the difference vector between the lens $i$ and
the source $k$; the angle $\varphi_{ik}$ is the polar angle of
$\vtheta_{ik}^{\rm ds}=\theta_{ik}^{\rm ds}\,\e^{\i\varphi_{ik}}$.

We then convert the stack $\Delta\gamma_{mn}$ of excess shear into a
map of the excess convergence. Owing to a sparse sampling of the shear
stack by discrete source positions around lens pairs, an additional
smoothing of this map is required. We apply this smoothing with a
kernel $K$ to $\Delta\gamma_{mn}$ before the conversion to the excess
mass map or the pair convergence map, namely by means of
\begin{equation}
  \label{eq:shearsmooth}
 \Delta\gamma_{mn}^{\rm K}=
 \frac{\sum\limits_{n^\prime,m^\prime=1}^{N_{\rm p},N_{\rm p}}
   W_{m^\prime n^\prime}\,K(m-m^\prime,n-n^\prime)\,\Delta\gamma_{m^\prime n^\prime}}
 {\sum\limits_{n^\prime,m^\prime=1}^{N_{\rm p},N_{\rm p}}W_{m^\prime n^\prime}\,K(m-m^\prime,n-n^\prime)}
\end{equation}
for the Gaussian kernel
\begin{equation}
 K(\delta m,\delta n)=
 \exp{\left(-\frac{1}{2}\frac{\delta m^2+\delta n^2}{\sigma_{\rm rms}^2}\right)}
\end{equation}
which has the kernel size $\sigma_{\rm rms}$ in units of our
grid-pixel size. Using the weights $W_{mn}$ for the smoothing ignores
grid pixel with no shear information and gives more weight to pixels
with a higher $W_{mn}$ in the average of neighbouring pixels. We use a
smoothing scale of \mbox{$\sigma_{\rm rms}=4$} for our maps.

In the last step, we apply the algorithm by
\citet{1993ApJ...404..441K} to $\Delta\gamma_{mn}^{\rm K}$ on the
grid, employing Fast-Fourier Transformations, to obtain the a smoothed
map
$\widehat{\Delta\kappa}^{\rm K}_{mn}=\Delta\kappa^{\rm
  K}_{mn}+\kappa_0$
of the excess convergence with a constant offset $\kappa_0$.  The real
part in the excess convergence contains the E-mode of the signal, and
the imaginary part is the B-mode. Applying the Kaiser-Squires
technique on a finite field produces systematic errors which typically
have the effect of increasing the signal towards the edges. We
therefore remove 50 pixel from the outer edges of the grid in the
final map. The inner cropped map has then the dimensions
$100\times100\,\rm pixel^2$. The constant offset $\kappa_0$ depends on
the details of the implementation of the Kaiser-Squire algorithm and
the number $N_{\rm p}$ of grid pixels. To have consistent maps in the
following, we assert that the average excess-convergence over the
cropped map has to vanish. We therefore subtract this average from the
final map.

\subsection{Galaxy-galaxy lensing and lens clustering}
\label{sect:gglomega}

The second-order statistics $\overline{\gamma}_{\rm t}(\vartheta)$ and
$\omega(\vartheta)$ are estimated from the data by the following
standard techniques. For the angular correlation function
$\omega(\vartheta)$ of the lens galaxies, we prepare a mock catalogue
with $n_{\rm r}$ uniform random positions within the unmasked region
of the survey. We then count the number
$DD(\vartheta;\Delta\vartheta)$ of lens pairs within the separation
bin
\mbox{$[\vartheta-\Delta\vartheta/2,\vartheta+\Delta\vartheta/2)$},
the number of random-galaxy pairs $RD(\vartheta;\Delta\vartheta)$, and
the number of random-random pairs $RR(\vartheta;\Delta\vartheta)$. For
the count rates, we consider all permutations of galaxy and mock
positions, which means the total number of $DD$, $RR$, and $DR$ for
all separations equals $n_{\rm d}(n_{\rm d}-1)$,
$n_{\rm r}(n_{\rm r}-1)$, and $n_{\rm d}n_{\rm r}$,
respectively. According to \cite{1993ApJ...412...64L}, we then
estimate (for $n_{\rm d},n_{\rm r}\gg1$)
\begin{equation}
  \label{eq:omegaest}
  \omega(\vartheta;\Delta\vartheta)=
  \frac{n_{\rm r}^2}{n_{\rm d}^2}\,
  \frac{DD(\vartheta;\Delta\vartheta)}{RR(\vartheta;\Delta\vartheta)}
  -2\,\frac{n_{\rm r}}{n_{\rm d}}\,
  \frac{DR(\vartheta;\Delta\vartheta)}{RR(\vartheta;\Delta\vartheta)}
  +1
\end{equation}
for the angular clustering of lenses at separation $\vartheta$.

To measure the mean tangential shear $\overline{\gamma}_{\rm t}$ and
the cross shear $\overline{\gamma}_\times$ within the separation bin
$[\vartheta-\Delta\vartheta/2,\vartheta+\Delta\vartheta/2)$, we apply
the estimator
\begin{equation}
  \label{eq:gglest}
  \overline{\gamma}_{\rm t}(\vartheta;\Delta\vartheta)+
  {\rm i}\,\overline{\gamma}_\times(\vartheta;\Delta\vartheta)=
  \frac{\sum\limits_{d,s=1}^{n_{\rm d},n_{\rm s}}
    \Delta_{ds}(\vartheta;\Delta\vartheta)\,w_s\,(-\ee^{-2{\rm i}\varphi_{ds}}\epsilon_s)}
  {\sum\limits_{d,s=1}^{n_{\rm d},n_{\rm s}}
    \Delta_{ds}(\vartheta;\Delta\vartheta)\,w_s}\;,
\end{equation}
where $\e^{-2\i\varphi_{ds}}=\vtheta_{ds}^\ast/\vtheta_{ds}^{}$ is the
phase factor of $\vtheta_{ds}=\vtheta^{\rm s}_s-\vtheta^{\rm d}_d$,
and $\Delta_{ds}(\vartheta;\Delta\vartheta)=1$ for
$\vartheta-\Delta\vartheta/2\le\theta_{ds}<\vartheta+\Delta\vartheta/2$
and $\Delta_{ds}(\vartheta;\Delta\vartheta)=0$ otherwise
\citep[e.g.,][]{2001PhR...340..291B}.

\begin{figure}
  \begin{center}
    \includegraphics[width=0.5\textwidth,trim=0cm 0cm 0cm 0cm,clip=true]{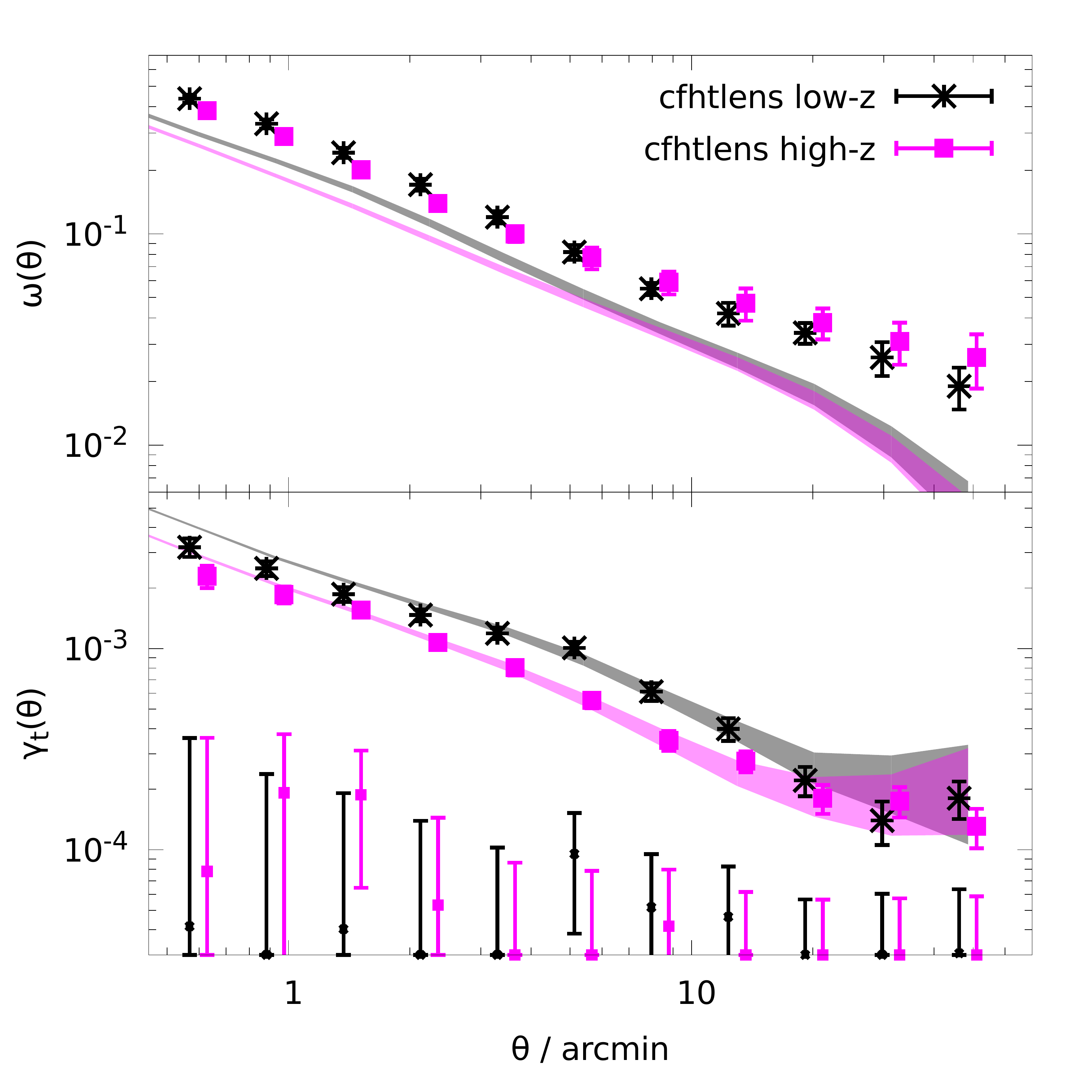}
  \end{center}
  \caption{\label{fig:omegaGGL} Angular clustering (top panel) and
    mean tangential shear (bottom panel) of our different lens
    samples. The large data points show the CFHTLenS measurements for
    the low-$z$ (stars) and high-$z$ sample (squares) with $1\sigma$
    error bars, and the coloured regions are $2\sigma$ predictions
    based on the H15 mocks. The grey regions are the predictions for
    the low-$z$ samples, the magenta regions predict the amplitude of
    the high-$z$ samples. The small data points with large errors bars
    at the bottom of the lower panel are the absolute values of the
    mean cross-shear for the CFHTLenS lenses.}
\end{figure}

Fig. \ref{fig:omegaGGL} is a summary of our measurements of
$\omega(\vartheta;\Delta\vartheta)$ and
$\overline{\gamma}_{\rm t}(\vartheta;\Delta\vartheta)$ for both the
CFHTLenS data and the H15 galaxy mocks. The measurements are
subdivided into the low-$z$ and high-$z$ redshift bins. We show the
tight constraints from the mocks as $2\sigma$ regions, and the
CFHTLenS measurements as large stars and square with $1\sigma$ error
bars (obtained by jackknife re-sampling, see
Sect. \ref{sect:jnerrors}). The small data points in the bottom panel
are the CFHTLenS cross-shear
$\overline{\gamma}_\times(\vartheta;\Delta\vartheta)$ of the lens
samples, which are plotted as absolute value in the logarithmic
plot. The mean of all cross-shear data points is consistent with
zero. Regarding a redshift dependence of the measurements, the low-$z$
data points are somewhat higher than the high-$z$ data points for both
statistics, a trend which is also predicted by the H15 model. The
model predictions for the tangential shear $\overline{\gamma}_{\rm t}$
are in very good agreement with the CFHTLenS data although the slope
for the low-$z$ profile is slightly shallower compared to H15 for
\mbox{$\vartheta\lesssim5\,\rm arcmin$}. A more detailed comparison of
the galaxy-galaxy-lensing signal to H15 can be found in S17.

The amplitude of the angular clustering $\omega(\vartheta)$ in the
model, on the other hand, is about 30\% lower than observed. This
might partly be explained by a distance distribution $p_{\rm d}(\chi)$
of lenses that is actually narrower in CFHTLenS than the assumed
distribution in the model. If so, galaxy-galaxy lensing
$\overline{\gamma}_{\rm t}(\vartheta)$ would be little affected as
long as the mean distance of lenses and sources is nevertheless
sufficiently accurate. The clustering amplitude $\omega(\vartheta)$,
however, would be affected more strongly because it depends on
$p^2_{\rm d}(\chi)$, although it is unlikely that a systematically
broadened $p_{\rm d}(\chi)$ in the model alone can fully explain the
observed discrepancy. This would require the RMS variance of
$p_{\rm d}(\chi)$ in the model to be biased high by as much as
40\%. We obtain this crude estimate by assuming a narrow top-hat shape
for $p_{\rm d}(\chi)$ with width $\Delta r$ and centre $r_{\rm c}$, as
in Equation (11) of \cite{2007A&A...473..711S}. In this case, the
amplitude of $\omega(\vartheta)$ scales with $(\Delta r)^{-1}$.

More likely, the disagreement between the clustering of H15 galaxies
and CFHTLenS galaxies reflects the current model uncertainties: in
particular, \cite{2017MNRAS.469.2626H} find a systematically too low
amplitude of the galaxy clustering at \mbox{$z=0.1$} in the stellar
mass range between \mbox{$9\le\log_{10}{(M_{\rm sm}/h^{-2}\,\msol)}<10$}
for the H15 model.  They find an amplitude too low by
\mbox{$\sim20-30\%$} at a projected separation between the galaxies of
a few \mbox{$100\,h^{-1}\,\rm kpc$} (see their Figure 13). This
comparison to galaxies from the Sloan Digital Sky Survey, albeit at
$z=0.1$ instead of \mbox{$0.2\lesssim z\lesssim0.6$}, is consistent
with our finding. The authors argue that the systematic error in the
galaxy clustering is related to the treatment of supernova feedback
and the gas reincorporation time in the model, affecting the
clustering and prevalence of low-mass galaxies. We also refer to the
recent work by \cite{2017MNRAS.466.2718C} for a thorough discussion on
the impact of SAM and simulation parameters on several observational
properties of galaxies. As to measurements of galaxy-galaxy-galaxy
lensing in this study, we point out that the estimator for the excess
mass map uses $1+\omega(\vartheta)$ rather than $\omega(\vartheta)$ at
\mbox{$\vartheta\approx1\,\rm arcmin$} for which the discrepancy is
around 7\%, and for the pair convergence map we do not use the angular
clustering of lenses at all.

\subsection{Convergence stack}
\label{sect:EstimatorSAM}

In a variant of the previous estimator for maps of the excess mass, we
stack the excess convergence $\Delta\overline{\kappa}_{\rm emm}$ or
$\Delta\overline{\kappa}$ around lens pairs in simulated data
directly. Let $\vec{\theta}^{\rm d}_i$ be the positions of $n_{\rm d}$
lens galaxies and $\kappa(\vec{\theta})$ a simulated grid of
convergence values. Similar to Sect. \ref{sect:EstimatorData}, we use
a shear-free affine transformation to map $\vec{\theta}$ positions
around a given pair $ij$ of lenses to the stacking frame with
coordinates $\vec{p}$. The estimation process proceeds in four steps:
(i) we stack the convergence around a set of selected lens pairs from
the separation bin
\mbox{$\vartheta-\Delta\vartheta/2\le|\vec{\theta}^{\rm
    d}_i-\vec{\theta}^{\rm d}_j|<\vartheta+\Delta\vartheta/2$}
to obtain $\overline{\kappa}_{\rm pair}$; (ii) we stack the
convergence around individual lenses to obtain
$\overline{\kappa}_{{\rm ind},1}+\overline{\kappa}_{{\rm ind},2}$ in
the stacking frame; (iii) we estimate $1+\omega$ averaged for the
distribution of lens-lens separations in the sample of selected lens
pairs; and (iv) we combine the steps (i) to (iii) to compute the
excess mass map
$\overline{\Delta\kappa}_{\rm emm}=(1+\omega)\overline{\kappa}_{\rm
  pair}-\overline{\kappa}_{{\rm ind},1}-\overline{\kappa}_{{\rm
    ind},2}$
with an estimate of $\omega$, which is averaged over the distribution
of lens-lens separations in the stack. As before, setting
\mbox{$\omega\equiv0$} in (iii) yields the pair convergence map.

For the steps (i) and (ii), we use square grids with $N_{\rm p}\times
N_{\rm p}$ pixels and coordinates $\vec{p}_{mn}=m+\i\,n$. The lens
positions are defined to be at the fixed location $\vec{p}^{\rm d}_1$
and $\vec{p}^{\rm d}_2$. Using the definitions \Ref{eq:affine} for the
parameters of the mapping between the $\vec{p}$-frame and the
$\vec{\theta}$-frame for a given lens pair $ij$, we obtain
\begin{equation}
  \vec{\theta}_{ij}^{mn}=A^{-1}_{ij}(\vec{p}_{mn}-B_{ij})
\end{equation}
for the position $\vec{\theta}$ in the convergence grid that
corresponds to the stack position $\vec{p}_{mn}$. We then compute the
stack $\overline{\kappa}_{\rm pair}$ for the grid pixel $mn$ by the
average
\begin{equation}
  \overline{\kappa}_{\rm pair}^{mn}=
  \frac{\sum_{i,j=1}^{n_{\rm d}}\Delta_{ij}^{mn}\,\kappa(\vec{\theta}_{ij}^{mn})}
  {\sum_{i,j=1}^{n_{\rm d}}\Delta_{ij}^{mn}}\;,
\end{equation}
where $\Delta_{ij}^{mn}=1$ indicates if $\vec{\theta}_{ij}^{mn}$ is
inside the $\kappa$-grid or $\Delta_{ij}^{mn}=0$ otherwise. In case of
$\Delta_{ij}^{mn}=1$, we choose the grid point in
$\kappa(\vec{\theta})$ that is closest to $\vec{\theta}_{ij}^{mn}$.

To obtain a map of the average convergence around individual lenses in
step (ii), we have to factor in that the convergence around pairs in
the stack is differently scaled for any new lens pair in the
stack. Therefore, to obtain
$\overline{\kappa}_{\rm ind,1}+\overline{\kappa}_{\rm ind,2}$ for a
distribution of scale parameters $|A_{ij}|$ we apply the following
technique.  For each lens galaxy at $\vtheta^{\rm d}_i$, we randomly
pick a position
$\vtheta^{\rm rnd}_i = \vtheta^{\rm d}_i+\delta\theta_i\,\ee^{\ii
  \phi_i} $
for an `imaginary' uncorrelated lens, where $\phi_i$ defines a
uniformly random orientation $\phi_i\in[0,2\pi)$ and $\delta\theta_i$
is a random separation from the $i$th lens. For $\delta\theta_i$, we
randomly pick a separation
$|\vec{\theta}^{\rm d}_k-\vec{\theta}^{\rm d}_l|$ from a list of all
separations of selected lens pairs in the data. This assures that (ii)
consistently uses the same distribution of rescalings $|A_{ij}|$ that
are applied in step (i). Since the positions of imaginary lenses are
uncorrelated with both $\kappa(\vec{\theta})$ and the lens positions,
a stack around lens-imaginary pairs yields (on average)
$\overline{\kappa}_{\rm ind,1}$ if the lens is mapped to
$\vec{p}^{\rm d}_1$ and the imaginary lens mapped to
$\vec{p}^{\rm d}_2$; we obtain $\overline{\kappa}_{\rm ind,2}$ if we
swap the lens positions in the stack. Therefore, by adding the maps
for both cases we obtain the stack
$\overline{\kappa}_{\rm rnd}=\overline{\kappa}_{\rm
  ind,1}+\overline{\kappa}_{\rm ind,2}$, which justifies the estimator
\begin{equation}
  \overline{\kappa}_{\rm rnd}^{mn}=
  \frac{\sum_{i=1}^{n_{\rm d}}
    \Delta^{mn}_{ij}\kappa(\vec{\theta}_{ij}^{mn})}
  {\sum_{i=1}^{n_{\rm d}}\Delta^{mn}_{ij}}\
  +
  \frac{\sum_{i=1}^{n_{\rm d}}
    \Delta^{mn}_{ji}\kappa(\vec{\theta}_{ji}^{mn})}
  {\sum_{i=1}^{n_{\rm d}}\Delta^{mn}_{ji}}\;,
\end{equation}
where $\Delta^{mn}_{ij}$ and $\vec{\theta}_{ij}^{mn}$ are defined as
before with the exception that we use
$\vec{\theta}^{\rm d}_j\equiv\vec{\theta}^{\rm rnd}_i$ for the
position of the second lens in the lens pair $ij$.

Finally, we combine the information from the previous steps to compute
the excess mass around the lens pairs by
\begin{equation}
  \overline{\Delta\kappa}_{\rm emm}^{mn}=
  \Big[1+\omega(\vartheta;\Delta\vartheta)\Big]\,
  \overline{\kappa}_{\rm pair}^{mn}-\overline{\kappa}_{\rm rnd}^{mn}\;.
\end{equation}
For a consistent comparison with maps obtained by shear stacking,
Sect. \ref{sect:EstimatorData}, we smooth the excess mass map with the
same kernel as in Eq. \Ref{eq:shearsmooth},
\begin{equation}
  \overline{\Delta\kappa}_{\rm emm,K}^{mn}=
  \frac{\sum\limits_{n^\prime,m^\prime=1}^{N_{\rm p},N_{\rm p}}
    K(m-m^\prime,n-n^\prime)\,\overline{\Delta\kappa}_{\rm emm}^{m^\prime n^\prime}}
  {\sum\limits_{n^\prime,m^\prime=1}^{N_{\rm p},N_{\rm p}}K(m-m^\prime,n-n^\prime)}\;,
\end{equation}
we apply the same cropping, and we subtract the average of
$\overline{\Delta\kappa}_{\rm emm,K}^{mn}$ from the cropped map.

\subsection{Combining measurements}
\label{sect:combine}

Our data consist of \mbox{$i=1\ldots n_{\rm f}$} separate fields: four
fields W1 to W4 for the CFHTLenS data and 64 fields for the synthetic
data. Separated fields means here that we ignore contributions from
galaxy pairs or triples where not all galaxies are inside the same
field. For a combined measurement, we apply the estimators described
in the previous sections for each field individually and then average
them as described in the following. Our strategy for performing
measurements of galaxy-galaxy-galaxy lensing with CFHTLenS data is an
improvement in comparison to \citet{2013MNRAS.430.2476S}. In that
work, measurements in $n_{\rm f}=129$ individual pointings were
performed and combined afterwards for a final result.  Here, using the
continuous fields W1-4, each consisting of many adjacent pointings,
allows us to include also galaxy tuples with galaxies from different
pointings. We find that this new strategy can enhance the overall S/N
in the CFHTLenS maps moderately by 10-30 per cent, depending on the
lens samples and their redshift binning.

For a combined estimate of the mean tangential shear
$\overline{\gamma}_{\rm t}(\vartheta)$, we imagine the application of
Eq. \Ref{eq:gglest} to a merged catalogue of all fields where
positions between any pair of galaxies from separate fields are larger
than the considered range of $\vartheta$.  For this merged catalogue,
let $n_{\rm d}^i$ and $n_{\rm s}^i$ be the number of lenses and
sources inside field $i$, and
\mbox{$\overline{\gamma}_{\rm t}^i+\i\overline{\gamma}_\times^i$} the
estimator \Ref{eq:gglest} for galaxies in field $i$ only. We then
split the sums over lens-source pairs in \Ref{eq:gglest} for the
merged catalogue into additive contributions from each field to obtain
\begin{equation}
  \overline{\gamma}_{\rm
    t}(\vartheta;\Delta\vartheta)+
  \i\overline{\gamma}_\times(\vartheta;\Delta\vartheta)=
  \frac{\sum_{i=1}^{n_{\rm f}}W^i\,\Big(\overline{\gamma}_{\rm
      t}^i(\vartheta;\Delta\vartheta)+
    \i\,\overline{\gamma}_\times^i(\vartheta;\Delta\vartheta)\Big)}
  {\sum_{i=1}^{n_{\rm f}}W^i}
\end{equation}
with the field weights
\begin{equation}
  W^i=
  \sum_{s,d=1}^{n_{\rm d}^i,n_{\rm s}^i}\Delta^i_{ds}(\vartheta;\Delta\vartheta)\,w_s^i
\end{equation}
for the combined estimate.  Here $w_s^i$ are the statistical weights
of sources $s$ in field $i$, and the flag
$\Delta^i_{ds}(\vartheta;\Delta\vartheta)$ applies to positions in
field $i$ only.

For the combined estimate of $1+\omega$, we have to determine the
count rates $DD$, $DR$, and $RR$ in the merged catalogue. To this end,
we cannot simply add the count rates of all individual fields. Instead
we have to pay attention to the field variations of numbers
$n_{\rm d}^i$ of observed galaxies and of unclustered random galaxies
in the merged survey. For example, the number of random positions
inside field $i$ should depend on the effective area of the field. To
quantify this, let $p_i$ be the probability that a random position of
unclustered galaxies is inside field $i$, and $n_{\rm r}$ is the total
number of random points for all fields; the distribution of mocks in
field $i$ complies with the selection function in field $i$. Only if
all fields have the same effective area (or selection function in
general), we will find \mbox{$p_i=p_j=1/n_{\rm f}$} for
\mbox{$i\ne j$}. This, for example, is exactly the case for the 64
fields in our simulated data, and approximately for the 129 CFHTLenS
pointings that make up the fields W1-4. For a measurement of the
angular clustering of lenses in CFHTLenS, we combine the count rates
inside the individual pointings, hence here $n_{\rm f}=129$; for all
other correlation functions we perform measurements inside the large
fields W1-4.  Now, counting the total number of random-random pairs in
the merged catalogue we find
\begin{multline}
  \frac{RR(\vartheta;\Delta\vartheta)}{n_{\rm r}^2}=\\
  \frac{1}{n_{\rm r}^2}\,\sum_{i=1}^{n_{\rm
      f}}RR^i(\vartheta;\Delta\vartheta) =\sum_{i=1}^{n_{\rm
      f}}p_i^2\,\frac{RR^i(\vartheta;\Delta\vartheta)}{p_i^2n_{\rm
      r}^2} =:\sum_{i=1}^{n_{\rm
      f}}p_i^2\,rr^i(\vartheta;\Delta\vartheta)\;,
\end{multline}
where \mbox{$rr^i:=RR^i/(p_i^2n_{\rm r}^2)$} is the count rate $RR^i$
in field $i$ normalised with the total number $(p_in_{\rm r})^2$ of
random pairs in this field. Conveniently, the value of $rr^i$ does not
depend, on average, on the absolute number of mock positions that we
actually in the individual measurement of field $i$. Therefore, for
the combined result of normalised counts $rr=RR/n_{\rm r}^2$, we take
the average of all individual $rr^i$ weighted with $p_i^2$. Similarly,
we obtain for the normalised count rate $dr=DR/(n_{\rm d}n_{\rm r})$
of $DR$ pairs in the merged catalogue
\begin{equation}
  dr(\vartheta;\Delta\vartheta):=
  \sum_{i=1}^{n_{\rm f}}\frac{p_i\,n_{\rm d}^i}{n_{\rm d}}\,
  \frac{DR^i(\vartheta;\Delta\vartheta)}
  {n_{\rm d}^ip_in_{\rm r}}=
  \sum_{i=1}^{n_{\rm f}}p_i\,f_i\,dr^i(\vartheta;\Delta\vartheta)\;,
\end{equation}
where $f_i=n_{\rm d}^i/n^{}_{\rm d}$ is the fraction of galaxies in
field $i$, and $dr^i$ refers to the normalised rate in field $i$
which, as before, does not depend on the number of mock positions
used. Consequently, we compute the combined $dr$ by taking a weighted
sum of individual $dr^i$. The normalised count $dd=DD/n_{\rm d}^2$ of
$DD$ pairs is
\begin{equation}
  dd(\vartheta;\Delta\vartheta):=
  \sum_{i=1}^{n_{\rm f}}f^2_i\,dd^i(\vartheta;\Delta\vartheta)\;.
\end{equation}
In summary, we compute the count rates $(dd^i,dr^i,rr^i)$ for each
separate field $i$ and then perform the previous weighted sums for the
combined rates $(dd,dr,rr)$. By means of the estimator
\Ref{eq:omegaest}, we then get
\begin{equation}
  \omega(\vartheta;\Delta\vartheta)=
  \frac{dd(\vartheta;\Delta\vartheta)}{rr(\vartheta;\Delta\vartheta)}-
  2\,\frac{dr(\vartheta;\Delta\vartheta)}{rr(\vartheta;\Delta\vartheta)}
  +1
\end{equation}
for the combined clustering amplitude of all fields.

With regard to combining measurements of the shear stacks from a set
of $n_{\rm f}$ separate fields we do the following. We apply the
estimator \Ref{eq:excessshear} with identical grid parameters to each
field $i$ for $\Delta\gamma_{mn}^i$ and the statistical weights
$W_{mn}^i$. Therein, we constantly use the clustering amplitude
$1+\omega$ as estimated once from the merged catalogue and the
measurement of the tangential shear $\overline{\gamma}_{\rm t}^i$ in
field $i$. This is consistent with the technique in
\citet{2008A&A...479..655S} but a slight variation compared to
\cite{2013MNRAS.430.2476S} where $1+\omega$ is measured for each
CFHTLenS pointing individually. We then combine the measurements of
the excess shear inside the fields into
\begin{equation}
  \Delta\gamma_{mn}=
  \frac{\sum_{i=1}^{n_{\rm f}}W^i_{mn}\,\Delta\gamma^i_{mn}}
  {\sum_{i=1}^{n_{\rm f}}W^i_{mn}}\;.
\end{equation}
Finally, we apply the smoothing \Ref{eq:shearsmooth} to the combined
$\Delta\gamma_{mn}$ and perform the remaining steps in
Sect. \ref{sect:EstimatorData} for a combined map of the excess mass.

For a combined measurement of the excess mass map with the
convergence-stack technique in Sect. \ref{sect:EstimatorSAM}, we just
take the equally weighted average of all grids
$\overline{\kappa}^{mn}_{\rm emm}$ obtained from the individual
$n_{\rm f}=64$ fields. An optimised weighting scheme is not necessary
here because this simplistic approach already produces maps with
negligible statistical noise for our simulated data.

\subsection{Statistical errors}
\label{sect:jnerrors}

For an estimate of the statistical error in the CFHTLenS maps of the
excess mass, we perform the jackknife technique \citep[see
e.g.,][]{knight1999mathematical}. The basic idea is as follows. Let
$G(d)$ be an estimator for some quantity $G$ based on the complete
data set $d$. In our case, $G(d)$ is our estimator for the smoothed
excess mass map (or pair convergence map)
$\overline{\kappa}^{\rm K}_{mn}$ at the grid pixel $mn$, and $d$
comprises the merged catalogue of all separate fields.  For the
jackknife estimator, we split the complete data
$d=d_1\cup d_2\cup\ldots\cup d_{n_{\rm jn}}$ into $n_{\rm jn}$
disjoint samples, and we compute a sample
$\{G_{-i}:i=1\ldots n_{\rm jn}\}$ of estimates
$G_{-i}:=G(d\,\backslash\,d_i)$ based on $d$ without the subset
$d_i$. We define with
\begin{equation}
  \sigma^2(G)=\frac{n_{\rm jn}-1}{n_{\rm jn}}\sum_{i=1}^{n_{\rm jn}}
  \left(G_\circ-G_{-i}\right)^2\;,
\end{equation}
for
\begin{equation}
  G_\circ=\frac{1}{n_{\rm jn}}\sum_{i=1}^{n_{\rm jn}}G_{-i}\;,
\end{equation}
the estimator of the jackknife error-variance of $G(d)$. For the
following maps, we compute the jackknife error
$\sigma(\overline{\kappa}^{\rm K}_{mn})$ for every pixel value
$\overline{\kappa}^{\rm K}_{mn}$ and quote
$\overline{\kappa}^{\rm K}_{mn}/\sigma(\overline{\kappa}^{\rm
  K}_{mn})$ for the S/N of a pixel value.

We expect that positional shot-noise and shape noise of the sources as
well as cosmic variance are the dominating contributors of statistical
noise in our measurement; see, for example,
\citet{2005A&A...442...69K} for a discussion of statistical noise in
related lensing correlation-functions. For the jackknife scheme, we
remove individual pointings $d_i$ from the merged catalogue; each
CFHTLenS pointing has a square geometry and $1\times1\,\rm deg^2$
area. Since these pointings are significantly larger than our maps
with typical angular scale of a few arcmin, we expect a sensible
estimate for errors owing to cosmic variance at these scales
\citep{2016arXiv160708679S}.

\section{Results}
\label{sect:results}

\begin{figure}
\centering
    \includegraphics[width=0.5\textwidth,trim=0cm 0cm 0cm 0cm,clip=true]{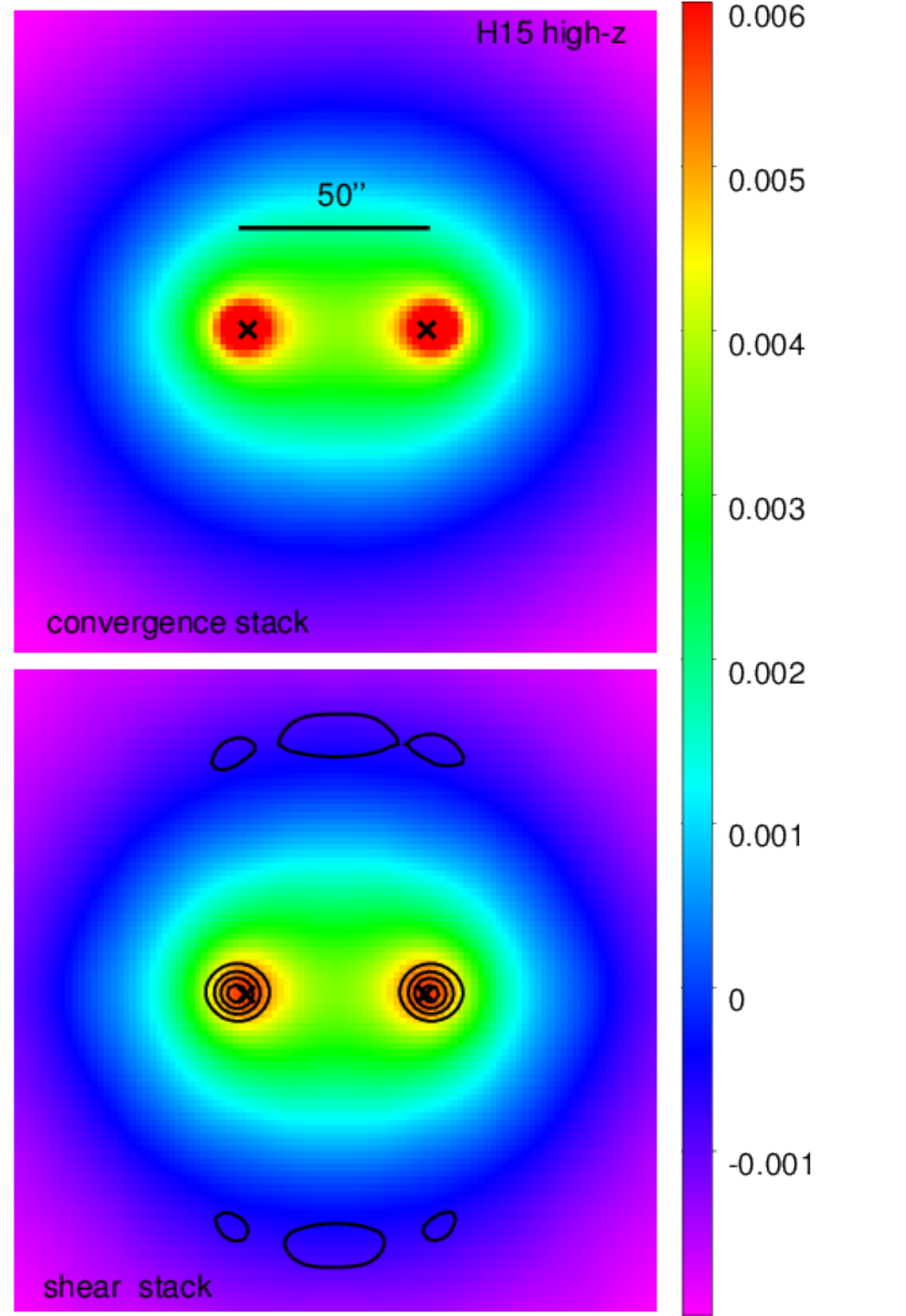}
    \caption{\label{fig:sep1EMMtest} Verification test with simulated
      data: A comparison of the E-mode excess mass
      $\overline{\kappa}_{\rm emm}$ (intensity scale) obtained from
      the $1024\,\rm deg^2$ mock data with two different method. The
      lens pairs are selected from the separation interval
      close-$\theta$ and redshift bin high-$z$
      (\mbox{$\bar{z}_{\rm d}\approx0.52$}). The lens positions inside
      the map are indicated by crosses. The shape and sampling noise
      of the sources are negligible here. \ita{Top}: Explicit
      stacking of the convergence on a grid; only applied to the
      simulated data. \ita{Bottom}: Reconstruction with a shear stack
      as applied to our CFHTLenS data. The contours show the relative
      difference
      \mbox{$(\overline{\Delta\kappa}_{\rm
          emm,2}-\overline{\Delta\kappa}_{\rm
          emm,1})/\overline{\Delta\kappa}_{\rm emm,1}$}
      between the two methods $\overline{\Delta\kappa}_{\rm emm,1}$
      (top) and $\overline{\Delta\kappa}_{\rm emm,2}$ (bottom) in
      steps of $10\%,20\%,\,\rm etc.$ for regions where the signal
      magnitude is above $5\times10^{-4}$.}
\end{figure}

In this section, we present maps of the excess mass
$\overline{\Delta\kappa}_{\rm emm}$ and pair convergence
$\overline{\Delta\kappa}$ for CFHTLenS lenses and for H15 galaxies in
our synthetic data. We compute the maps for two photo-$z$ bins and two
angular separations of lenses on the sky: the separation bins
\mbox{$40^\pprime\le\theta_{12}<60^\pprime$} (`close-$\theta$') and
\mbox{$60^\pprime\le\theta_{12}<80^\pprime$} (`wide-$\theta$'); and
the two redshift bins \mbox{$0.2\le z_{\rm ph}<0.44$} (`low-$z$') and
\mbox{$0.44\le z_{\rm ph}<0.6$} (`high-$z$'). We only report E-mode
maps in this section; the corresponding B-mode maps can be found in
the Appendix. The B-modes are consistent with a vanishing
signal. Moreover, we study the impact of a redshift slicing of lenses
as means to reduce the fraction of chance pairs in a shear stack.

\subsection{Code verification}
\label{sect:KappaShearMaps}

For Fig. \ref{fig:sep1EMMtest}, we compare the reconstruction of the
excess mass with synthetic data for two different methods: shear
stacking (Sect. \ref{sect:EstimatorData}) and convergence stacking
(Sect. \ref{sect:EstimatorSAM}). For convenience, in the high-$z$ bin
we show only the reconstructions for close-$\theta$ galaxies. The
relative deviations of maps for other samples, including the maps of
the pair convergence, are comparable to the single case shown here.
The bottom panel employs, for $\overline{\Delta\kappa}_{\rm emm,2}$,
the shear-stacking that we apply to the CFHTLenS data. The top panel
displays the map $\overline{\Delta\kappa}_{\rm emm,1}$ using direct
convergence stacking. Each of the 64 fields contains $2\times10^4$
sources with no intrinsic shape noise, to reduce the noise in the maps
for this comparison. Inside the panels, we indicate the lens positions
$\vec{p}_1^{\rm d}$ and $\vec{p}_2^{\rm d}$ by black crosses. We
overall find an excellent agreement for both approaches, with relative
differences usually around five per cent or less, wherever the signal
is larger than $5\times10^{-4}$. However, the differences grow larger
close to the lens positions inside the map, as indicated by the
contours of
\mbox{$(\overline{\Delta\kappa}_{\rm
    emm,2}-\overline{\Delta\kappa}_{\rm
    emm,1})/\overline{\Delta\kappa}_{\rm emm,1}$},
shown with the error levels in increments of 10\%. Nevertheless, the
errors are below 20\% except within a couple of pixels separation from
the lens positions. Presumably this error is owing to pixellation and
binning of the correlation function $\overline{\gamma}_{\rm t}$. The
10\% error contours at the edges of the map are likely just numerical
noise, which becomes relevant for values of
$\overline{\Delta\kappa}_{\rm emm,1}$ that are close to zero; the
convergence stacking randomly picks a separation to an imaginary lens
to subtract second-order correlations from the map. We reiterate that
shear stacking is insensitive to constant offsets in $\kappa$ so that
this level of agreement here is only valid for a consistent definition
of $\kappa_0$.

\begin{figure*}
  \centering
  \includegraphics[width=0.46\textwidth,trim={0cm 0cm 0cm 0cm},clip=true]{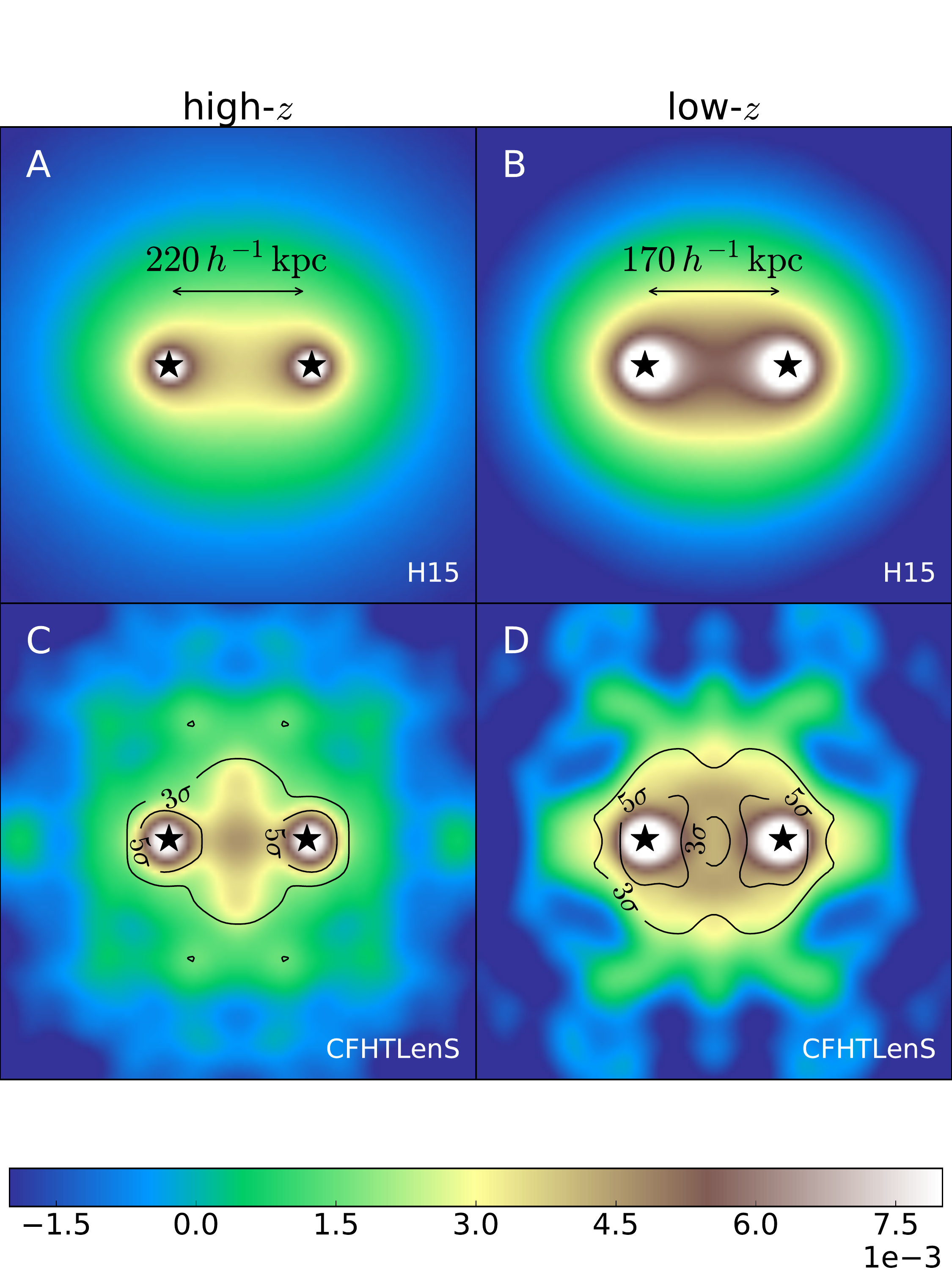}
  \includegraphics[width=0.46\textwidth,trim={0cm 0cm 0cm 0cm},clip=true]{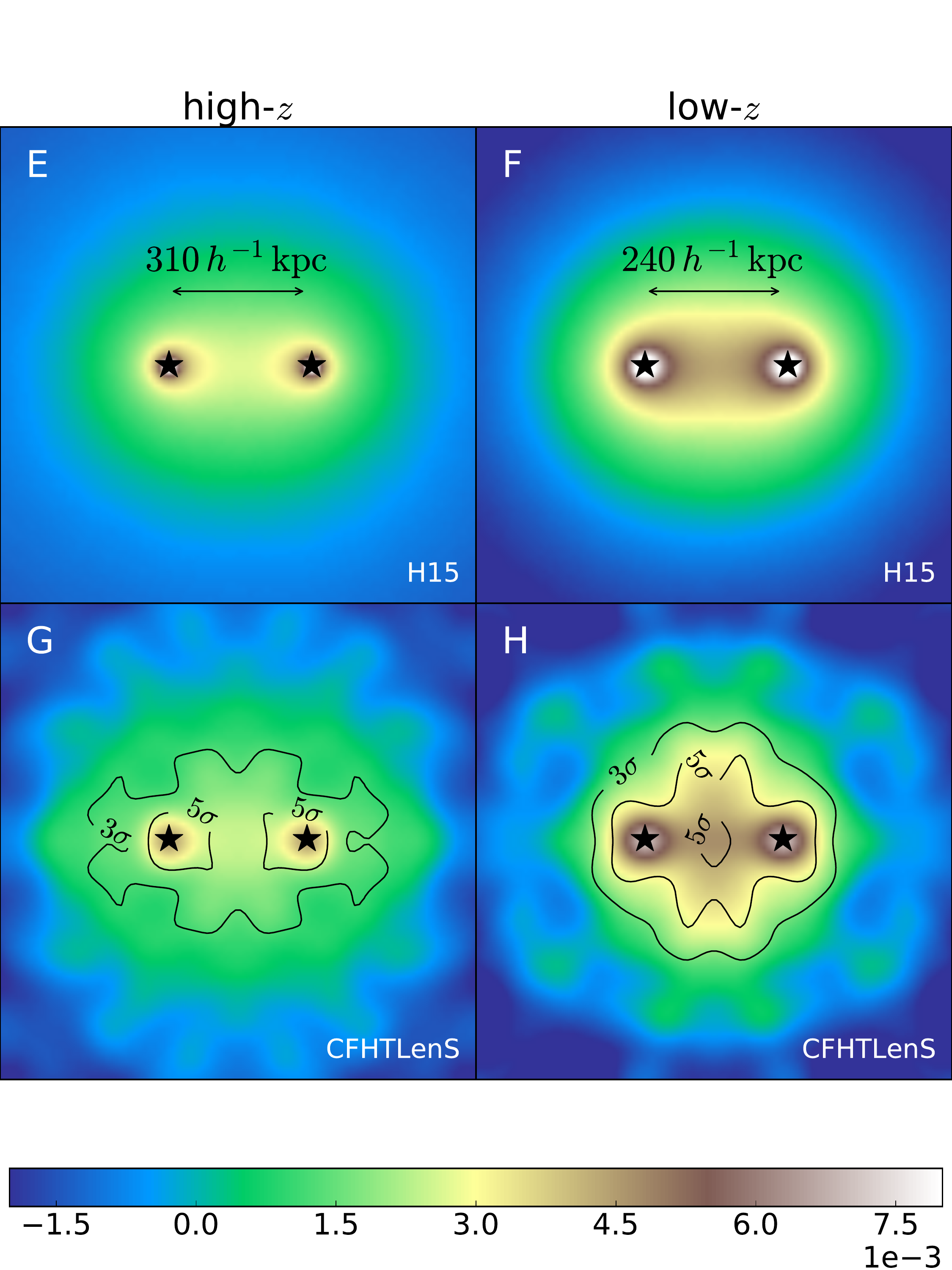}
  \\
  \includegraphics[width=0.46\textwidth,trim={0cm 0cm 0cm 0cm},clip=true]{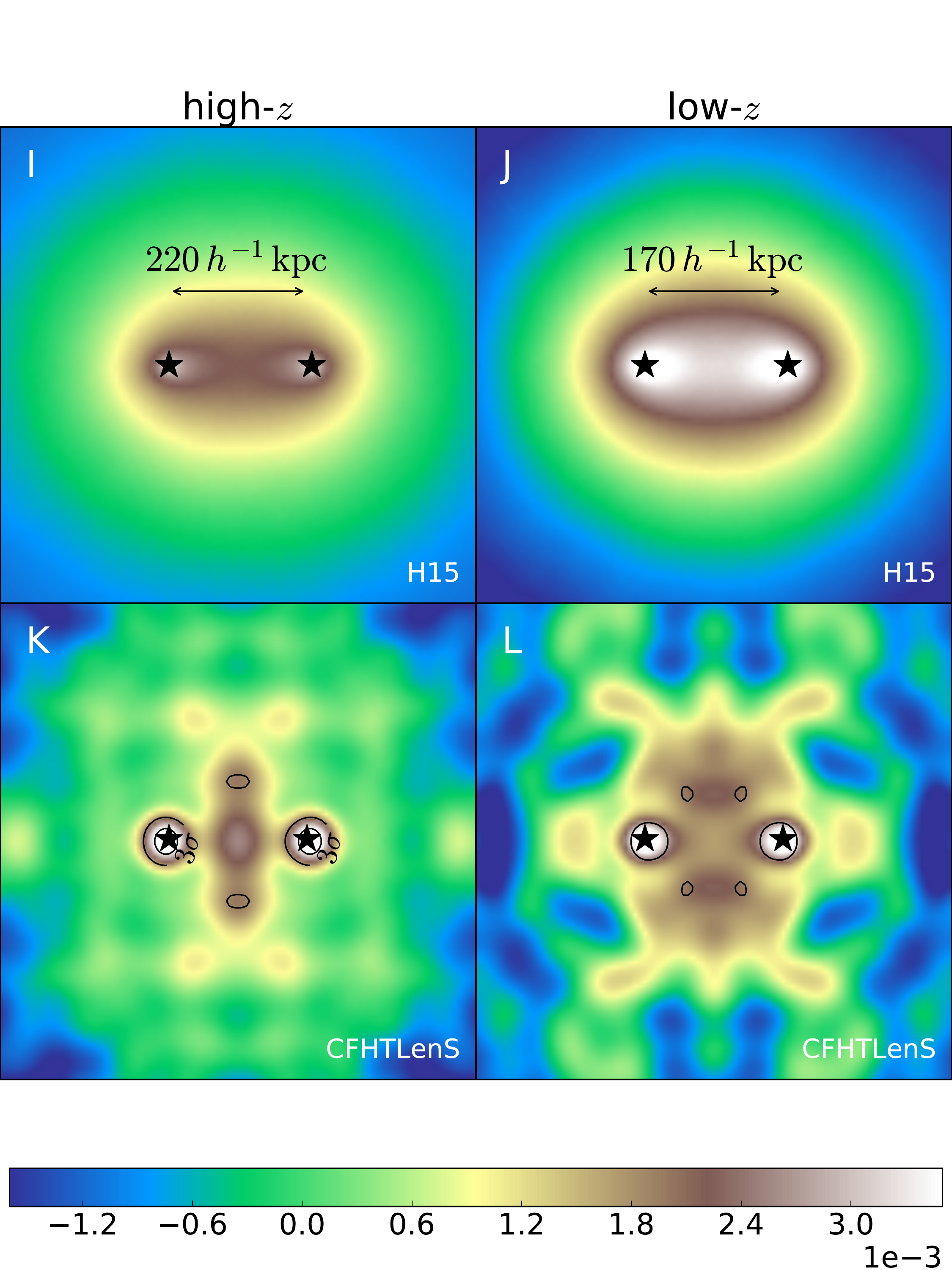}
  \includegraphics[width=0.46\textwidth,trim={0cm 0cm 0cm 0cm},clip=true]{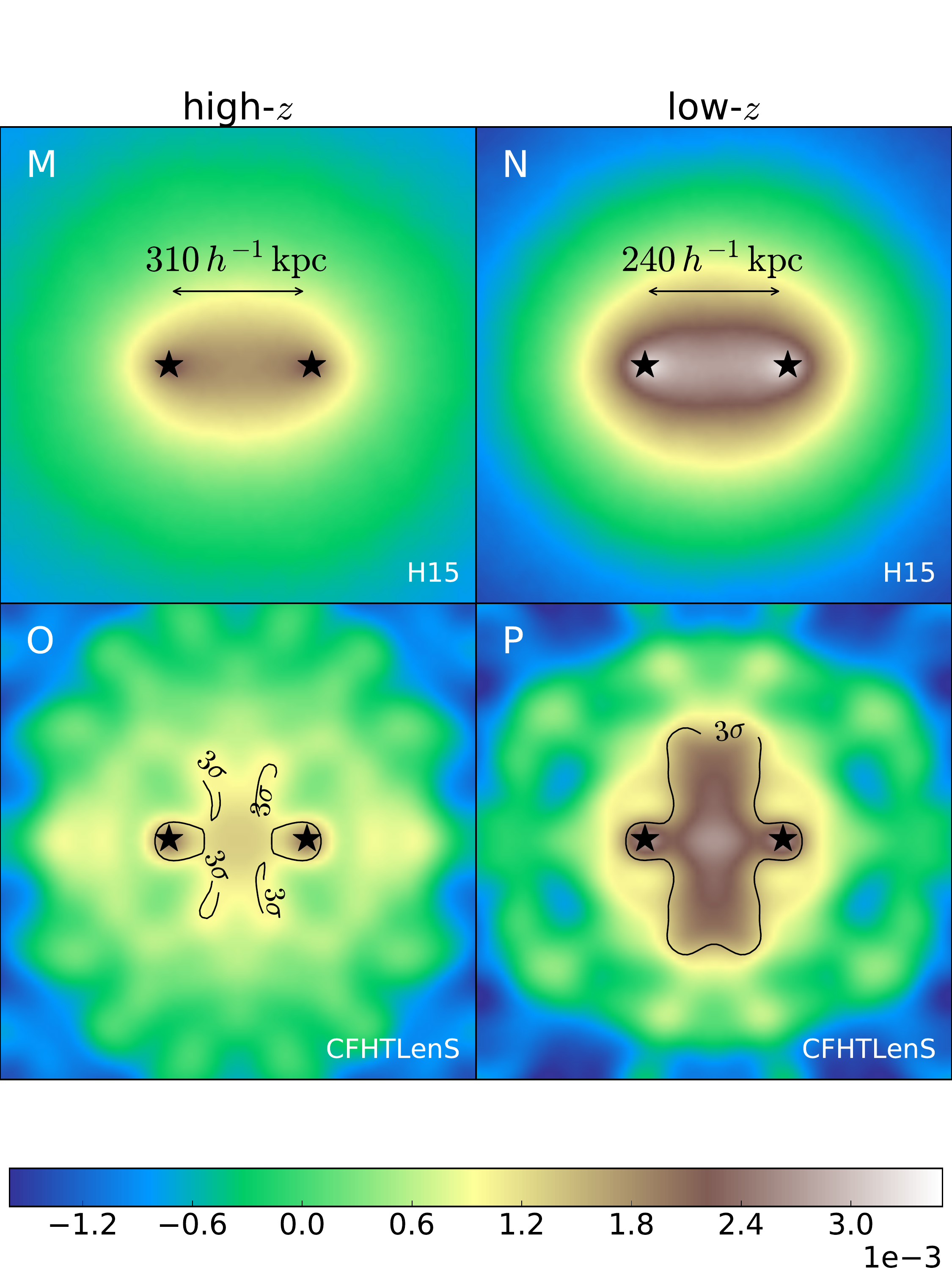}
  \caption{\label{fig:comparison} Comparison of the E-mode excess mass
    (panels A-H) and the pair convergence (panels I-P) between
    CFHTLenS data (from shear stacks) and the H15 model (from
    convergence stacks). The maps are arranged in bins of lens
    redshifts (columns) and projected angular separations of
    lenses. The mean angular separations for close-$\theta$ and
    wide-$\theta$ lenses are converted into projected distances and
    quoted inside the panels. The lens positions are indicated by
    stars. The contour lines indicate the significance levels
    $3\sigma$ and $5\sigma$ for the CFHTLenS measurements.}
\end{figure*}

\subsection{Excess mass maps}
\label{sect:EMMmocksCFHTLenS}

The panels A to H in Fig. \ref{fig:comparison} show the estimated
excess mass $\overline{\Delta\kappa}_{\rm emm}$, as measured in
CFHTLenS (bottom panels C, D, G, H), in comparison to the H15 model
predictions (top panels A, B, E, F). The samples are split by column
into the low-$z$ and high-$z$ redshift bins; see the labels at the
top. The mean redshifts and the angular separations of lenses inside
the bins correspond to a range of projected distances between
\mbox{$\approx170-300\,h^{-1}\,\rm kpc$}; the two lens positions are
indicated by stars.  We quote the distance scale inside the
panels. The measurements in CFHTLenS are subject to strong statistical
noise due to positional shot-noise and shape noise of the sources.
Based on a jackknife resampling of the data, the contour lines
encircle areas of $3\sigma$ and $5\sigma$ significance. Clearly, we
find a better than $3\sigma$ detection only in the central part of the
map, close to and between the lens positions. In comparison, cosmic
variance and shot noise by lenses are the only noise components
present in the panels for the H15 data. Similar to the H15 model
predictions for the aperture statistics in S17, statistical noise is
negligible here so that we do not show S/N contours.

The B-mode of the excess mass in the top panels of
Fig. \ref{fig:bmodes} (Appendix) are indicators of systematic
errors. To enhance the significance of the indicators, we also combine
the B-mode maps for low-$z$ and high-$z$ in this figure (third
column). The consistency of the indicators with random noise supports
that systematic errors are negligible inside the maps, at least those
that can be detected by B-modes.

Apart from noise in the CFHTLenS maps, we find a good agreement with
the predictions for H15 galaxies: a strong concentration of excess
mass close to the lens positions and a drop of the signal by about
$\Delta\kappa_{\rm emm}=4\times10^{-3}$ from the centre to the outer
regions of the maps. This drop corresponds to a change in the excess
surface-mass density of
$\Delta\kappa_{\rm emm}\times\overline{\Sigma}_{\rm
  crit}\approx17\,h\,\msol\,\rm pc^{-2}$
for the fiducial value of $\overline{\Sigma}_{\rm crit}$ in
Eq. \Ref{eq:sigmacrit}. We also observe a tentative indication of a
morphological difference between CFHTLenS data and H15, in particular
for the panels A vs. C and F vs. H: the excess mass has a bulge-like
distribution in vertical direction which is absent in the simulations;
the distribution in the simulated maps is more concentrated along the
line connecting the lenses inside the map.

Intrigued by this tentative feature in the CFHTLenS data, we produce
for a quantitative analysis significance maps of the residuals between
CFHTLenS and H15, which are the two panels in
Fig. \ref{fig:residuals}. For these two maps, we first compute shear
stacks from a synthetic lens and shear catalogue, using exactly the
same binning parameters and lens selections as for the CFHTLenS shear
stacks. Then we subtract the model shear-stacks from the CFHTLenS
shear-stacks, and we then proceed as described in the
Sects. \ref{sect:EstimatorData} and \ref{sect:jnerrors} to produce a
S/N map of the residual excess mass. The subtraction of shear steaks
avoids the problem of the unknown offset $\kappa_0$ in the maps. As we
are interested in the potential conflicts between H15 and CFHTLenS
data, we additionally combine in this process the (residual) shear
stacks of the low-$z$ and high-$z$ samples for the same separation
bin, giving us two maps instead of four: the left panel for
close-$\theta$ lenses and the right panel for wide-$\theta$ lenses. In
these maps, we find an agreement between CFHTLenS and H15 within
$3\sigma$ with the exceptions of four spots, close to the position of
the lenses inside the maps. Here the residuals become negative and
attain up to $3.5\sigma$ significance (inside the dashed
contours). Inside these spots, the CFHTLenS signal is thus lower than
the H15 excess mass, which gives rise to the appearance of a squeezed,
elongated bulge between the spot positions in the panels C, D, and H
of Fig. \ref{fig:comparison}.

\subsection{Pair convergence maps}
\label{sect:PairConvergenceMap}

In the panels I to P of Fig. \ref{fig:comparison}, we plot the
distribution of pair convergence $\overline{\Delta\kappa}$ for our
galaxy samples. Again, we have the H15 model predictions in the top
panels I, J, M, N, and the corresponding CFHTLenS measurements in the
bottom panels K, L, O, and P; the samples are split in redshift and in
angular separation of the lenses. We have pointed out in
Sect. \ref{sect:chancepairs} that the distribution of the pair
convergence is less affected by chance pairs. However, the
then-missing contribution by chance pairs in the maps has the effect
that the significance levels for the pair convergence are lower, and
the overall amplitude of the signal is also diminished by the fraction
of chance pairs in the lens sample. For instance, the significance of
the pair convergence between lenses is now just below $3\sigma$ for
close-$\theta$. In addition, the signal drop $\Delta\kappa$ from the
map centre to the outer regions is typically smaller for the pair
convergence, roughly by a factor of two, namely
$2\times10^{-3}\,\overline{\Sigma}_{\rm crit}\approx8\,h\,\msol\,\rm
pc^{-2}$.
Again, the B-mode maps in the lower part of Fig. \ref{fig:bmodes}
(Appendix) imply negligible systematic errors.

\begin{figure*}
  \begin{center}
    \includegraphics[width=0.9\textwidth,trim=0cm 0cm 0cm 0cm,clip=true]{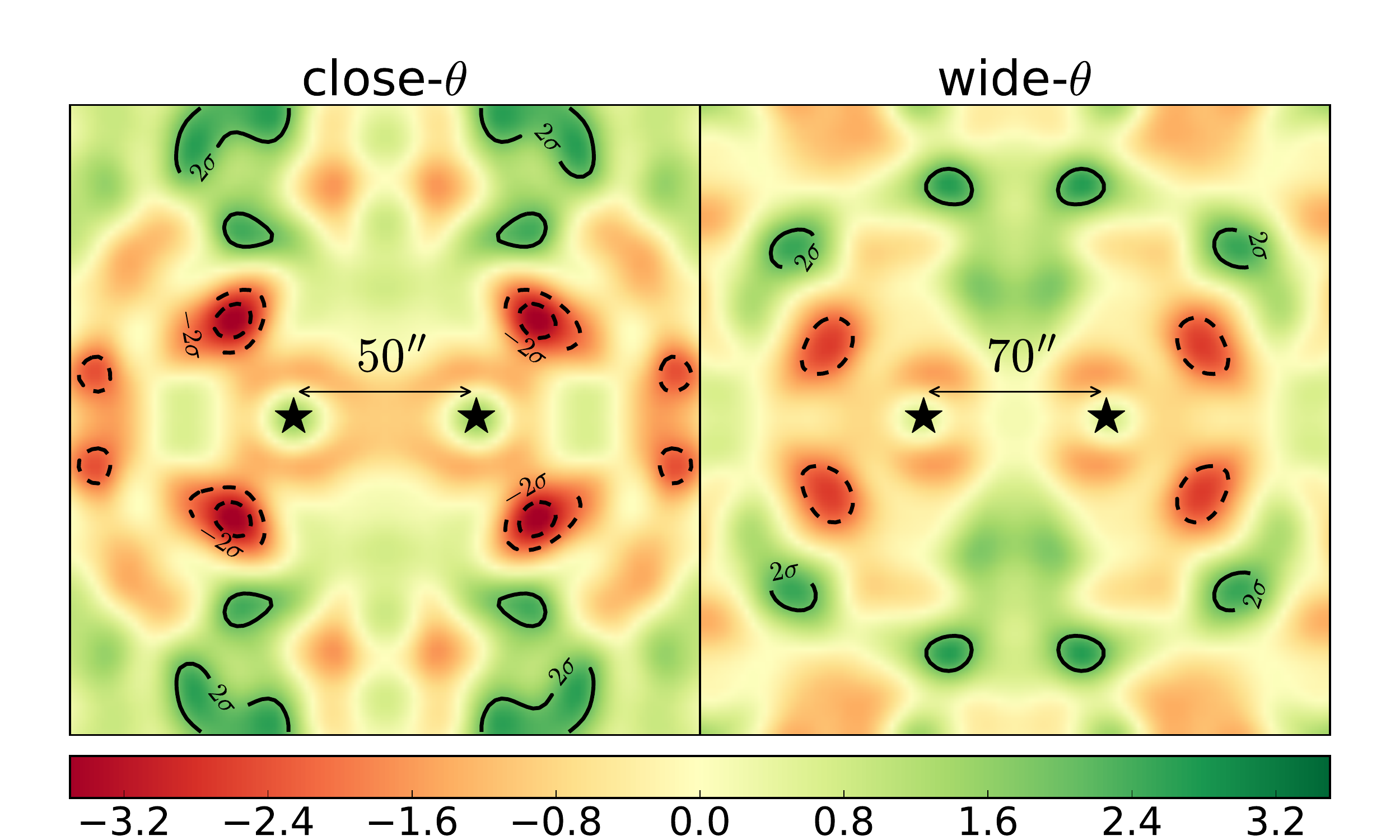}
  \end{center}
  \caption{\label{fig:residuals} Signal-to-noise ratio of residuals in
    the CFHTLenS excess mass after subtraction of the shear stacks
    from the H15 model predictions (E-mode).  To increase the
    significance in both maps, we combine the residual signals from
    the low-$z$ and high-$z$ samples for the same angular
    separation. The contours indicate regions of $2\sigma$ or
    $3\sigma$ significance; dashed lines are for negative
    residuals. \ita{Left panel:} Model residuals for the combined
    close-$\theta$ separation bin. \ita{Right panel:} Model residuals
    for the wide-$\theta$ separation bin.}
\end{figure*}

Moreover, the difference between the signals at the lens positions
(stars) and the centre of the maps is smaller than for the excess
mass. For the pair convergence of the mock data, we have no pronounced
peak at the lens positions but just a diffuse halo in which the lenses
are embedded. Although this is similar for the CFHTLenS wide-$\theta$
samples (panels O and P), this is not the case for the close-$\theta$
samples (panels K and L). This may, in part, be related to the 10-20
per cent inaccuracy of the shear stacking that we find within a few
pixels of the lens positions; see
Sect. \ref{sect:KappaShearMaps}. More prominently, and different to
the H15 prediction, there is a bulge in the distribution of the pair
convergence between the galaxies; see in particular the `gummy-bear'
like shape in panel P. Similar to the excess mass maps
$\overline{\Delta\kappa}_{\rm emm}$, this is produced by a relative
suppression of signal at the four corners around the lens pair and may
be a tentative indication of a disagreement between the H15 prediction
and CFHTLenS galaxies. The significance and pattern of residuals is
similar to Fig. \ref{fig:residuals} and not shown here for this
reason.

\subsection{Fraction of correlated lens pairs}
\label{sect:correlatedpairs}

It is interesting to estimate the number of true pairs in the shear
stacks of CFHTLenS that are used for Fig. \ref{fig:comparison}. There
is, however, no clear distinction between true pairs and chance pairs
in reality, in contrast to the idealised discussion in
Sect. \ref{sect:chancepairs}. Nevertheless, we can define a
statistical measure for the fraction of correlated pairs in a stack by
looking at the redshift difference $|z_i-z_j|$ of the lenses $i$ and
$j$ in a pair: galaxies have a typical correlation length of
$r_0\sim5\,h^{-1}\rm Mpc$ and quickly decorrelate for separations
greater than several $r_0$, making them essentially uncorrelated
chance pairs. Therefore, as practical measure for the level
$\hat{p}_{\rm tp}$ of correlated pairs in a shear stack, we count the
number of pairs in a sample and angular separation bin, and we count
the expected number $n_{\rm rnd}$ of pairs for lenses with same
positions on the sky but uncorrelated redshifts. Based on this we use
\mbox{$\hat{p}_{\rm tp}:=(n_{\rm pair}-n_{\rm rnd})/n_{\rm pair}$} as
a measure for the fraction of excess pairs; uncorrelated lenses, such
as chance pairs, would exhibit no excess. In principle, in future
studies we could increase the fraction of correlated pairs in a stack
by rejecting pairs with \mbox{$|z_i-z_j|>\Delta z$}, provided precise
redshift estimators are available ($r_0$ corresponds to
\mbox{$\Delta z\approx10^{-3}$} for lenses that are on the same
line-of-sight). Realistically, a rejection below several $10^{-3}$ for
$\Delta z$ is not sensible because peculiar velocities of the
galaxies, especially for those in galaxy clusters, spread out the
radial distribution of correlated galaxy pairs.
 
To quantify $\hat{p}_{\rm tp}$ for our CFHTLenS lens pairs and for
additional hypothetical $\Delta z$-cuts applied to CFHTLenS, we use
the H15 lenses (which emulate CFHTLenS but have exact redshifts) in
the following way.
\begin{itemize}
\item For different $\Delta z$, we count in H15 the number
  $n_{\rm pair}$ of pairs with \mbox{$|z_i-z_j|\le\Delta z$} in the
  close-$\theta$ or wide-$\theta$ redshift bin.
\item Then we randomly reassign the redshifts of the lenses in the
  sample by bootstrapping (with replacement), and we again count the
  number $n_{\rm rnd}$ of pairs with \mbox{$|z_i-z_j|\le\Delta z$}
  but in the randomised sample. For the reassignment of redshifts,
  we only allow as random redshift for a lens $k$ the redshift of
  another lens $l$ with similar stellar mass. This means, the
  difference in stellar mass has to be
  $|\log_{10}({\rm sm}_k/{\rm sm}_l)|\le0.2\,\rm dex$ for the
  stellar masses ${\rm sm}_k$ and ${\rm sm}_l$. We apply this
  restriction in the bootstrapping because the redshift distribution
  of lenses is probably slightly dependent on brightness and hence
  stellar mass, which has to be accounted for in the randomised
  sample to avoid a bias in $n_{\rm rnd}$.
\end{itemize}
\begin{figure}
  \begin{center}
    \includegraphics[width=0.5\textwidth,trim=0cm 0cm 0cm 0cm,clip=true]{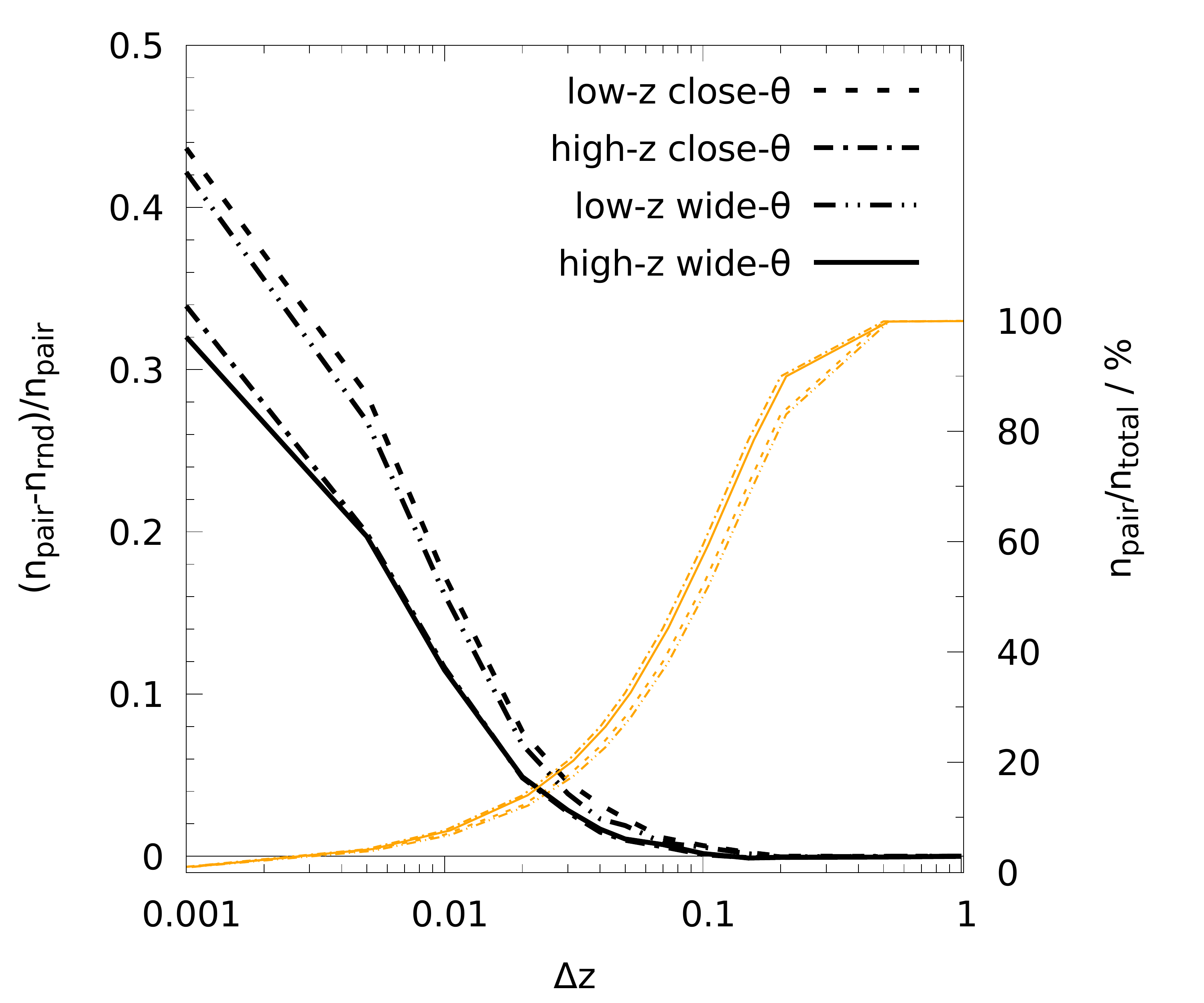}
  \end{center}
  \caption{\label{fig:H15truepairs} Plot of the estimated fraction
    $(n_{\rm pair}-n_{\rm rnd})/n_{\rm pair}$ of correlated lens pairs
    in a shear stack (thick lines and $y$-axis on the left-hand side)
    for lenses that have redshift difference smaller than $\Delta z$
    ($x$-axis). We obtain the plotted values for low-$z$ or high-$z$
    from H15 by counting the number of pairs $n_{\rm pair}$ and the
    pairs $n_{\rm rnd}$ with randomised redshifts for the angular
    separations close-$\theta$ or wide-$\theta$.  In addition, the
    thin orange lines and the right-hand $y$-axis show for each
    separation bin the number of lens pairs $n_{\rm pair}$ relative to
    the total number of pairs, i.e., the $n_{\rm pair}$ for
    $\Delta z\gg1$.}
\end{figure}
The thick black lines in Fig. \ref{fig:H15truepairs} and the left-hand
axis quote our results for $\hat{p}_{\rm tp}$ for lenses in the two
redshift bins and the two angular separation bins. This shows that
using lens pairs in low-$z$ or high-$z$ with a maximum redshift
difference of $\Delta z=10^{-3}$ increases $\hat{p}_{\rm tp}$ to about
$32-45\%$, whereas ignoring the lens redshifts ($\Delta z\gg1$), as we
do in the CFHTLenS analysis, reduces $\hat{p}_{\rm tp}$ to below
$0.1\%$. On the other hand, excluding lens pairs with redshift
differences greater than $\Delta z$ reduces the number of pairs in a
shear stack by the values that are given by the thin orange lines and
the right-hand $y$-axis. The different line styles belong to the
various photo-$z$ and separation bins. We find that a choice of
$\Delta z=0.1$ reduces the number of pairs to about 50\% of the total
number of pairs in low-$z$ or high-$z$, whereas $\Delta z\sim10^{-3}$
cuts this number down to below a few per cent.  For completeness, we
also give the mean redshift difference of lens pairs in H15, which is
typically for both separation bins and photo-$z$ bins
$\ave{|z_i-z_j|}=4.0\times10^{-3},2.4\times10^{-2},6.7\times10^{-2}$
for $\Delta z=0.01,0.05,0.1$. In CFHTLenS and our photo-$z$ and
separation bins, where we apply no $\Delta z$ cut, this value is
$\ave{|z_i-z_j|}=0.115$.

\subsection{Enriching the levels of correlated lens pairs}

\begin{figure*}
  \centering
  \includegraphics[width=0.95\textwidth,trim={0cm 0cm 0cm 0cm},clip=true]{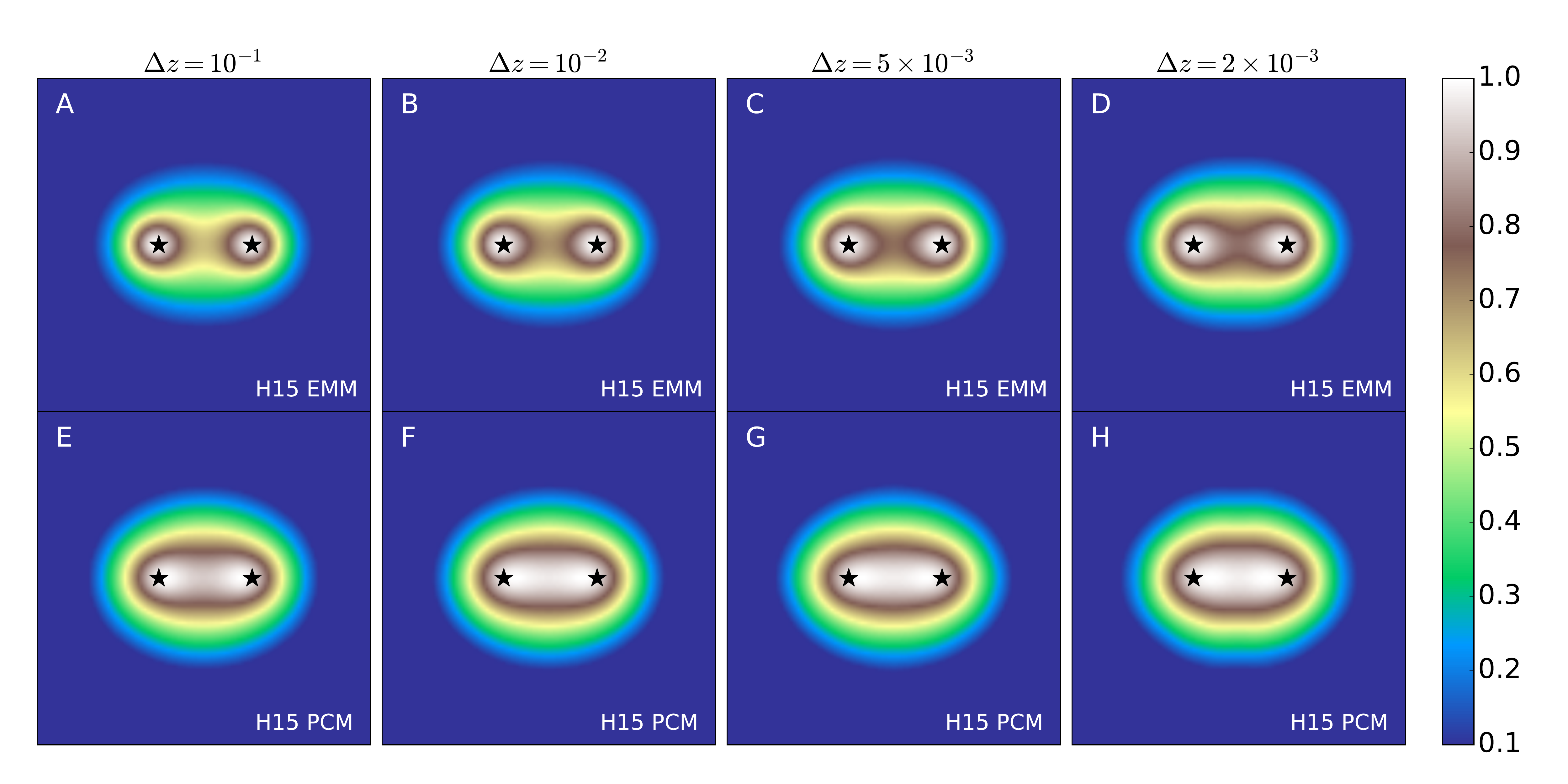}
  \caption{\label{fig:zbinning} Simulated change in the excess mass
    (EMM; panels A to D) and pair convergence (PCM; panels E to H)
    around lens pairs when enriching the level of correlated pairs by
    further subdividing the lens samples into slices with width
    $\Delta z= 10^{-1},10^{-2},5\times10^{-3},2\times10^{-3}$ (left to
    right). The shear stacks of the slices are combined for the final
    maps inside the panels. To highlight the relative differences
    between the maps, all maps are normalised to the maximum signal in
    the map; the intensity scale varies over the range 10\% to 100\%
    of the maximum. The maps use the H15 high-$z$ sample and pairs,
    indicated by the stars, that fall into the close-$\theta$
    separation bin.}
\end{figure*}

The previous section implies that the level of correlated lens pairs
in a shear stack may be increased by rejecting pairs that are well
separated in redshift. Here we briefly study the impact of this
rejection on the maps by using the synthetic data. For this purpose,
we focus on the high-$z$ sample and H15 lenses from the close-$\theta$
separation bin only; the low-$z$ lenses or the wide-$\theta$
separation bin have a qualitatively similar behaviour. We subdivide
the full high-$z$ lens sample into redshift slices of constant width
$\Delta z$ such that lens pairs in each slice have at most a
separation of $\Delta z$ in redshift. For each slice, we stack the
shear field around the lenses separately and combine all separate
shear stacks into one final shear stack later on, as outlined in
Sect. \ref{sect:combine}. In particular, for each individual stack we
measure and utilise the mean tangential shear
$\overline{\gamma}_{\rm t}(\vartheta)$ and angular clustering
$\omega(\vartheta)$ of the lenses in this slice only. We note that
this way of rejecting well-separated pairs is wasteful, especially for
a very fine slicing, because pairs which have
\mbox{$|z_i-z_j|\le\Delta z$} but where each lens resides in a
different slice are also rejected. While this is acceptable for our
synthetic data, which lacks shape noise, it is sub-optimal for
applications to noisy real data. We leave the development of a less
wasteful technique to future work.

Fig. \ref{fig:zbinning} is a display of the resulting simulated maps
for four different slicing parameters
$\Delta z=10^{-1},10^{-2},5\times10^{-3},2\times10^{-3}$ and redshifts
within the range $0.4\le z\le0.7$; the panels A to D are for the
excess mass, the panels E to H show the pairs convergence around an
average lens pair. We highlight the qualitative changes between the
maps when going to a finer slicing by normalising each map to its
maximum value. This hides the most prominent effect, namely that the
amplitude of the maximum signal in the excess mass maps increases from
\mbox{$\Delta\kappa_{\rm emm}/10^{-2}=1.8$} to $10.4$ from the
coarsest to the finest slicing. This amplitude increase is directly
related to the amplitude dependence of the aperture statistics
$\ave{{\cal N}^2M_{\rm ap}}$ on the variance of $p_{\rm d}(\chi)$
(Sect. 5.3 in S13).  The trend for the pair convergence is less clear:
the maximum signal progresses along
$\Delta\kappa/10^{-3}=5.3,\,8.1,\,6.2,\,3.2$ from $\Delta z=10^{-1}$
to $2\times10^{-3}$. The drop in the pair convergence amplitude below
$\Delta z\sim5\times10^{-3}$ despite an increasing $\hat{p}_{\rm tp}$
might indicate that we start to remove lens pairs from the sample that
carry a significant correlation signal in the shear stacks. This
means, the restrictive slicing might start to affect
$\overline{\kappa}_{\rm pair}(\vec{\vartheta}|\theta_{12})|_{\rm tp}$
in Eq. \Ref{eq:paircon2} of the simplified model.

Apart from the overall change in amplitude the change in the
normalised maps is minor for the excess mass (cf. panel A and D) and
invisible for the pair convergence. This observation is consistent
with the discussion in Sect. \ref{sect:chancepairs}, where we assume a
clear-cut dichotomy of correlated `true' and uncorrelated `chance'
pairs. In this simplified scenario, the normalised pair convergence is
unchanged when increasing the fraction of true pairs, whereas for the
excess mass the chance pairs add extra convergence close to the lens
positions in the map, weighted by the fraction of true pairs.  As the
fraction $\hat{p}_{\rm tp}$ increases from panel A to panel D, we
therefore anticipate a change in the normalised excess-mass map close
to the lens positions, which is indeed visible: we start to see a
merging of the halos around the lens positions to one common halo in
the excess mass map; similar to what can be seen for the pair
convergence.

\section{Discussion}
\label{sect:discussion}

Our work is the first direct comparison of the galaxy-galaxy-matter
correlation function, measured on real data at arc-minute scales, to
the predictions of a galaxy model. We demonstrate that shear-based
estimators directly and accurately measure the correlated surface-mass
density around physical galaxy pairs, by using a combination of
synthetic data, fine redshift slicing of lenses, and direct stacking
of convergence. We then apply this method to photo-$z$ binned lenses
in real data. Furthermore, we demonstrate that small distortions in
the excess mass maps, caused by nonphysical (chance) pairs near the
lens positions, can be suppressed by the newly introduced
pairs-convergence statistics. We discuss this in depth in the
following.

The results for the excess mass maps in the panels C, D, G, and H in
Fig. \ref{fig:comparison} show a clear $3\sigma-6\sigma$ detection
within the inner few $100\,h^{-1}\,\rm kpc$ around CFHTLenS
galaxy-pairs, and the signal has the expected qualitative behaviour
for changes in galaxy separation and redshift. Specifically, our
E-mode signal in the excess mass map decreases with lens separation at
a fixed redshift (as can be seen by comparing panels C to G and panels
D to H) and it decreases with redshift (as seen by comparing panels C
to D and panels G to H). Both trends are broadly expected, as the same
angular separation of galaxies on the sky at increasing redshift
corresponds to larger projected separation, and the three-point
correlation function of the matter density field decreases with
physical scale; galaxies are essentially tracers of the matter density
and therefore their three-point correlations have a similar
qualitative behaviour. The signal also varies with the lensing
efficiency as expressed by Eq. \Ref{eq:sigmacrit}. The relative
decrease of $\overline{\Sigma}_{\rm crit}$ between low-$z$
($\bar{z}_{\rm d}=0.35$) and high-$z$ ($\bar{z}_{\rm d}=0.52$) is
approximately $15\%$ and thus contributes to a signal decrease. An
additional signal decrease towards higher redshift may be given by
gravitational growth of structure which amplifies the non-Gaussianity
of the density fields with time. Finally, the consistency of explicit
convergence stacking and shear stacking with synthetic data in
Sect. \ref{sect:KappaShearMaps} shows that the CFHTLenS maps can be
interpreted as maps of excess convergence that is correlated with
galaxy pairs, in concordance with the theory presented in
Sect. \ref{sect:ExcessMass}. In addition, Fig. \ref{fig:zbinning}
underscores that the characteristic patterns in the maps originate
from physically close lens pairs that have
$\Delta z\lesssim5\times10^{-3}$, even if the majority of pairs are
chance pairs as in our CFHTLenS data (Fig. \ref{fig:H15truepairs}).

We have also introduced the pair convergence $\overline{\Delta\kappa}$
as alternative measure of the excess mass
$\overline{\Delta\kappa}_{\rm emm}$ to suppress the contamination by
chance (uncorrelated) pairs in the correlation statistics, and we
detect it with a significance of $3\sigma-4\sigma$ for our CFHTLenS
lenses. The maps are shown in panels K, L, O, and P in
Fig. \ref{fig:comparison}.  As to the motivation for an alternative
measure, our mapping technique of the excess mass stacks shear around
galaxy pairs, within a given separation interval on the sky. However,
the vast majority of selected pairs are actually well separated in
physical distance; they are uncorrelated chance pairs. In fact, in
each of the low-$z$ and high-$z$ samples fewer than 5\% of our lens
pairs have redshift differences of $\Delta z=5\times10^{-3}$ or less
(see Fig. \ref{fig:H15truepairs}). Fortunately, the main effect of the
chance pairs is to dilute the correlation signal and to slightly
distort the excess mass $\overline{\Delta\kappa}_{\rm emm}$ close to
the lens positions inside the map (see
Sect. \ref{sect:chancepairs}). It may be desired to remove this
distortion by rejecting chance pairs in the stack, which, however, is
difficult owing to the typical errors \mbox{$\sigma(z)\approx0.04$} of
the photometric-redshift estimations in CFHTLenS.  Therefore, we have
defined by Eq. \Ref{eq:pc} the slightly modified statistic of the pair
convergence $\overline{\Delta\kappa}$ which is not distorted by
uncorrelated pairs (but also has a somewhat different meaning).  On
the other hand, the (weakly distorted) excess mass
$\overline{\Delta\kappa}_{\rm emm}$ may be, after all, mathematically
preferable because it is only a function of the matter-galaxy
bispectrum and thus exactly vanishes for Gaussian density fields
\citep{2005A&A...432..783S}.

As expected from the recent work by S17, there is a good agreement
between the H15 predictions for the excess mass and our measurements
with CFHTLenS data, although we find some evidence for morphological
differences to the SAM predictions, especially for the pair
convergence. S17 measured, among other things: the $\cal G$-related
aperture statistics $\ave{{\cal N}^2M_{\rm ap}}(\theta_{\rm ap})$ for
angular scales $\theta_{\rm ap}$ between 1 and 10 arcmin, for their
galaxy samples sm1 to sm6 in both the low-$z$ and high-$z$ redshift
bins, and they find a good match between observations and the H15
model for all stellar-mass samples, reported in their Table 6.
Therefore, we expect that the H15 excess-mass maps should also agree
well with the measurements despite probing somewhat smaller angular
scales.  Indeed the H15 distribution of excess mass and its overall
amplitude is consistent with CFHTLenS, as can be seen from panels A to
H in the Fig. \ref{fig:comparison}. With regard to possible
differences between the H15 predictions and CFHTLenS, we notice a
bulge of excess mass in the vertical direction at the centre the
excess-mass maps. This can especially be seen in the panels H and P in
comparison to the models F and N, respectively, in
Fig. \ref{fig:comparison}. To quantify the differences, we produce two
maps of model residuals in Fig. \ref{fig:residuals} where we find good
agreement between model and CFHTLenS almost everywhere inside the maps
within $3\sigma$ confidence, with the exceptions of four spots of
negative residuals with approximately $3.5\sigma$ significance
(close-$\theta$ in the left panel) and $3\sigma$ significance
(wide-$\theta$ in the right panel). The corresponding map for the pair
convergence is similar and therefore not shown. The suppression of the
correlation signal at these spots produces the elongated bulge of
excess mass perpendicular to the orientation of the lens pair.

We have taken precautions to suppress spurious signals in the lensing
maps owing to intrinsic alignments of sources, and have performed
basic tests to confirm that the bulge feature is not induced by a
trivial systematic error. For overlapping distance distributions of
lenses and sources, we may find sources at lens distances. In this
case, correlations of intrinsic source ellipticities with the matter
density around lens pairs can add signal to our maps of the excess
mass or pair convergence. Although this effect is currently not well
studied, S13 argue that the alignment signal can be suppressed by
reducing the overlapping area $A$ of $p_{\rm d}(z)$ and
$p_{\rm s}(z)$, where \mbox{$A:=\int\d z\,a(z)$} is the integral over
\mbox{$a(z)=\min{\{p_{\rm d}(z),p_{\rm s}(z)\}}$}.  Using photometric
redshifts, our separation of lens and source distributions is not
perfect, but the overlap of the distributions is small: approximately
$A=4\%$ for low-$z$ and $A=12\%$ for high-$z$ (S13). In particular,
the bulge is still visible for panel H in Fig. \ref{fig:comparison}
despite the small $4\%$ overlap for low-$z$. Therefore, the bulge is
probably unrelated to intrinsic alignments. Furthermore, numerical
artefacts in the computer code that produce a bulge feature are also
unlikely as can be seen by the verification test in
Fig. \ref{fig:sep1EMMtest}. Here systematic errors are present close
to the lens positions only. We note that masking of data is not
included in the verification test, however this is not a plausible
cause of this effect: mask-related systematic errors should not have a
preferred map direction, unless the orientation of galaxy pairs at
\mbox{$z_{\rm d}\gtrsim0.2$} is correlated with the orientation of
mask features. As masking is mainly produced by satellite tracks,
stars, instrumental CCD effects, cosmic rays, or low-redshift galaxies
\citep{2013MNRAS.433.2545E}, this is unlikely. As another possible
systematic effect, we test if blending of galaxy images could bias the
shear estimates of CFHTLenS sources and affect the excess maps
\citep{2013MNRAS.429.2858M}. For the test, we have produced new maps
with same binning parameters as in the panels A to H in
Fig. \ref{fig:comparison}, although now rejecting sources within nine
arcsec of a lens galaxy for the shear stack. We find no significant
difference to the maps without rejection, and, in particular, the
bulge feature persists (not shown to save space). Finally, we stack
the shear patterns of all panels K, L, O, and P in
Fig. \ref{fig:comparison} to investigate a possible residual pattern
in the B-modes that is correlated with the prominent bulge feature in
the pair convergence maps.  We overlay the B-mode signal as black
contours in Fig. \ref{fig:bmodesall} on top of the combined
E-mode. The merged stacks of all lenses clearly show the bulge at the
centre, although the B-mode amplitude is typically below
$5\times10^{-4}$ near the bulge and thus small compared to the E-mode
signal. However, the B-mode signal is correlated with the bulge, and
extends to larger separations in the vertical direction at the
positions of the lenses (crosses) compared to its vertical extension
at the bulge location. This may equally be a coincidence or an
indication of a systematic effect for the bulge
appearance. Nonetheless, a clear interpretation of an alignment
between a residual B-mode signal and the E-mode bulge is difficult,
since we lack a model for systematic errors here that convincingly
connects both.

On the speculative side, if the bulge in the distribution of excess
mass is indeed a physical effect it could point to missing elements in
the dark-matter simulation or the SAM galaxy model (or the similar
SAMs in S17) used for this study. One conceivable element might be
the extra lensing signal caused by the intra-cluster medium (ICM) that
constitutes $10-15\%$ of the mass of matter halos in galaxy
clusters. On the one hand, if the ICM distribution is aligned with the
distribution of dark matter, there will be no qualitative change in
the distribution of excess mass, because additional gravitational
lensing by the ICM only rescales the dominating dark-matter signal. On
the other hand, a misalignment between the ICM and the dark-matter
distribution might produce a weak bulge feature if the ICM density is
increased perpendicular to the lens-lens axis. This could be tested
with cosmological simulations that include baryons, or analytically
with a halo model that includes misaligned mass-distributions
\citep[e.g.,][]{2002PhR...372....1C}. Similarly, we could imagine a
bulge feature being induced by a statistical misalignment between the
orientation of lens pairs and the orientation of their parent halo. As
an extreme example, one might consider a distribution of galaxies
distributed in the equatorial plane of a prolate halo: stacking the
surface matter-density of the halo around pairs of these galaxies
would result in an elongated excess signal perpendicular to the
orientation of lens pairs. This too could be tested with simulations
or a halo model by populating non-spherical dark-matter halos with a
misaligned galaxy distribution.

By selecting galaxy pairs with projected separations of around
$250\,h^{-1}\,\rm kpc$ our analysis probes the matter environment of
pairs inside galaxy groups and clusters. While this regime is of
particular interest to test the predictive power of galaxy models,
such as H15, future applications this probe may also be utilised to
analyse the filamentary large-scale distribution of matter around
physical pairs. To this end, future works should aim to map the pair
convergence around luminous red galaxies at $\sim10\,h^{-1}\rm Mpc$
separation, in a way that is similar to that presented in
\cite{2017arXiv170208485E} and \cite{2016MNRAS.457.2391C}, where
successful detections have been reported.

%
\section*{Acknowledgements}

We thank the anonymous referee for the thoughtful comments.  This work
has been supported by the Deutsche Forschungsgemeinschaft through the
project SI 1769/1-1 and through the Collaborative Research Center TR33
`The Dark Universe'. Patrick Simon also acknowledges support from the
German Federal Ministry for Economic Affairs and Energy (BMWi)
provided via DLR under project no. 50QE1103. Stefan Hilbert
acknowledges support by the DFG cluster of excellence ‘Origin and
Structure of the Universe’ (\url{www.universe-cluster.de}).

\bibliographystyle{aa}
\bibliography{BibFiles}

\appendix
\counterwithin{figure}{section}

\section{Additional figures}

\begin{landscape}
\begin{figure}
  \centering
  \vspace{2cm}
  \includegraphics[scale=0.6,clip=false,angle=-90]{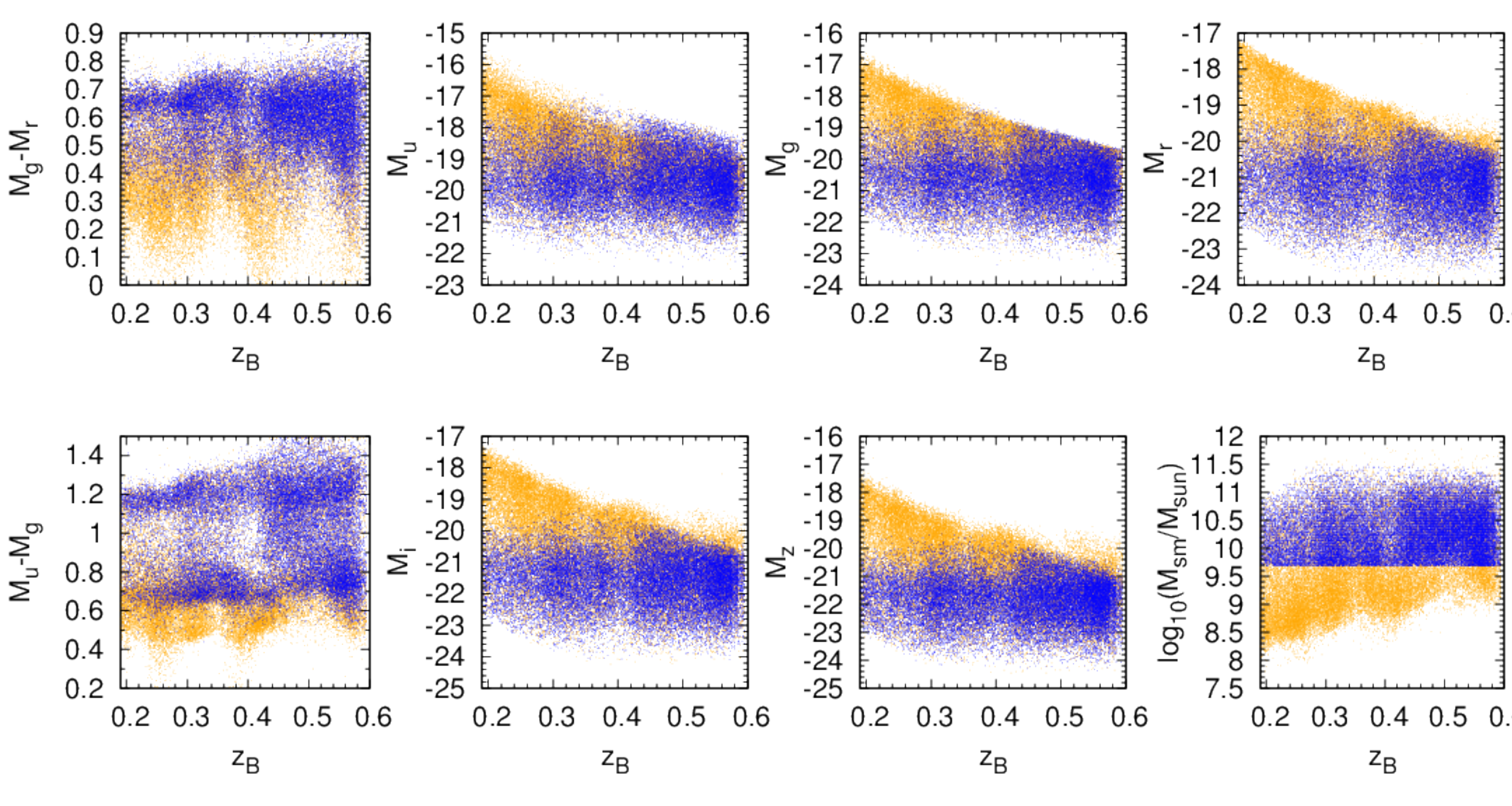}
  \caption{\label{fig:completeness} Properties of the targeted sample
    of lens galaxies in the lensing analysis. Shown are scatter plots
    of absolute rest-frame
    $u^\ast g^\prime r^\prime i^\prime z^\prime$-magnitudes, colours,
    and stellar masses versus the photometric redshift $z_{\rm ph}$
    (BPZ) of the CFHTLenS galaxies with flux limit
    \mbox{$i^\prime\le22.5$} in the stellar-mass interval
    \mbox{$5\times10^9\le M_{\rm sm}<3.2\times10^{11}\,\Msolar$} (blue
    dots). The brighter orange dots show the scatter for lenses for
    all stellar masses which are clearly flux limited at higher
    redshifts $z_{\rm ph}$.  The data points are from galaxies in the
    field W1 only; other fields look similar.}
\end{figure}
\end{landscape}

\begin{figure*}
  \begin{center}
    \includegraphics[width=0.85\textwidth,trim=0cm 0cm 0cm 0cm,clip=true]{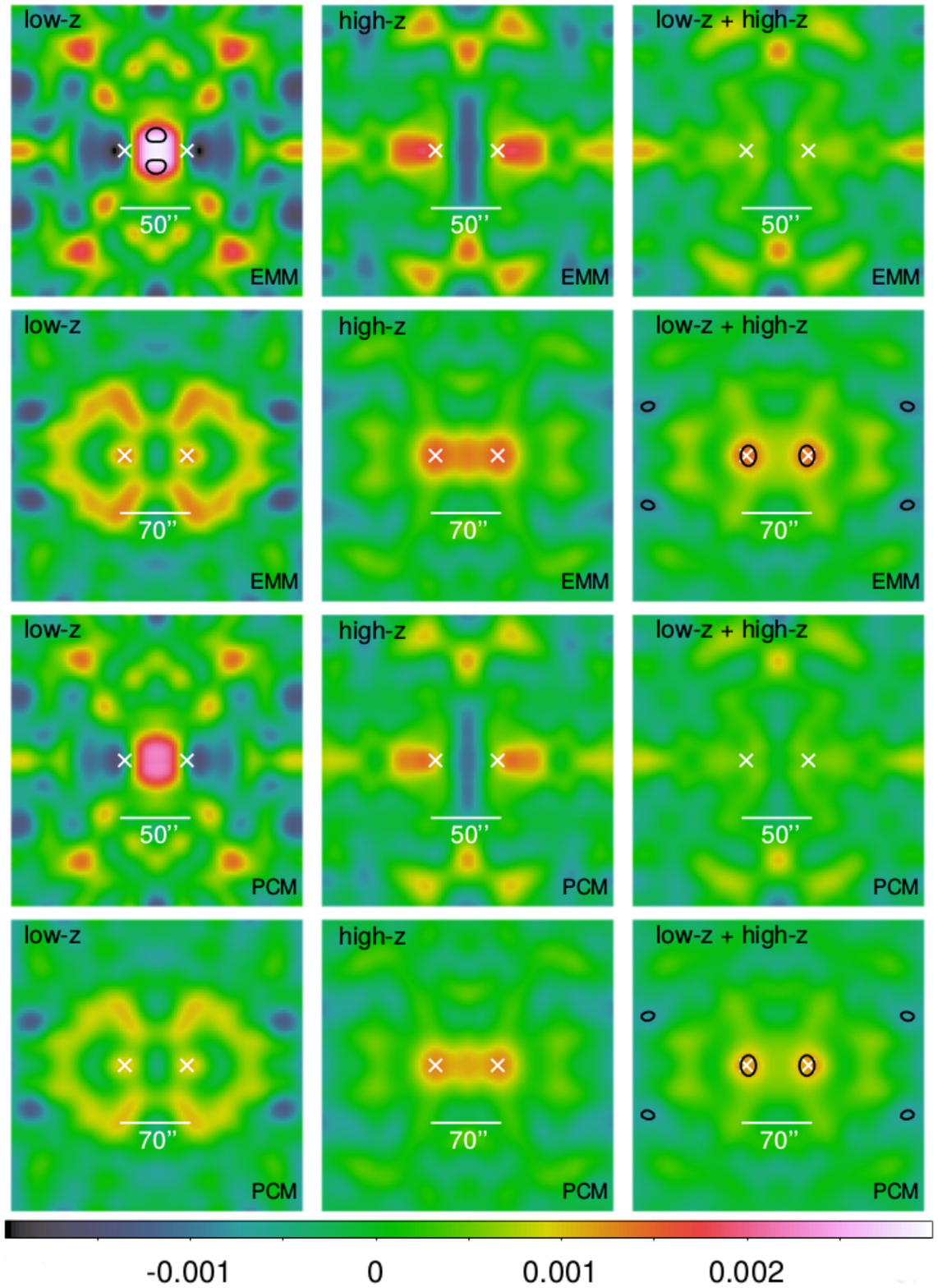}
  \end{center}
  \caption{\label{fig:bmodes} B-mode maps of the excess mass around
    CFHTLenS galaxy pairs for the excess mass (top panels) and the
    pair convergence (bottom panels). The angular scale and redshift
    selections are indicated inside the panels. The crosses show the
    lens positions inside the map. The contours indicate regions with
    significance greater or equal $3\sigma$.}
\end{figure*}

\begin{figure*}
  \begin{center}
    \includegraphics[width=0.6\textwidth,trim=0cm 0cm 0cm 0cm,clip=true]{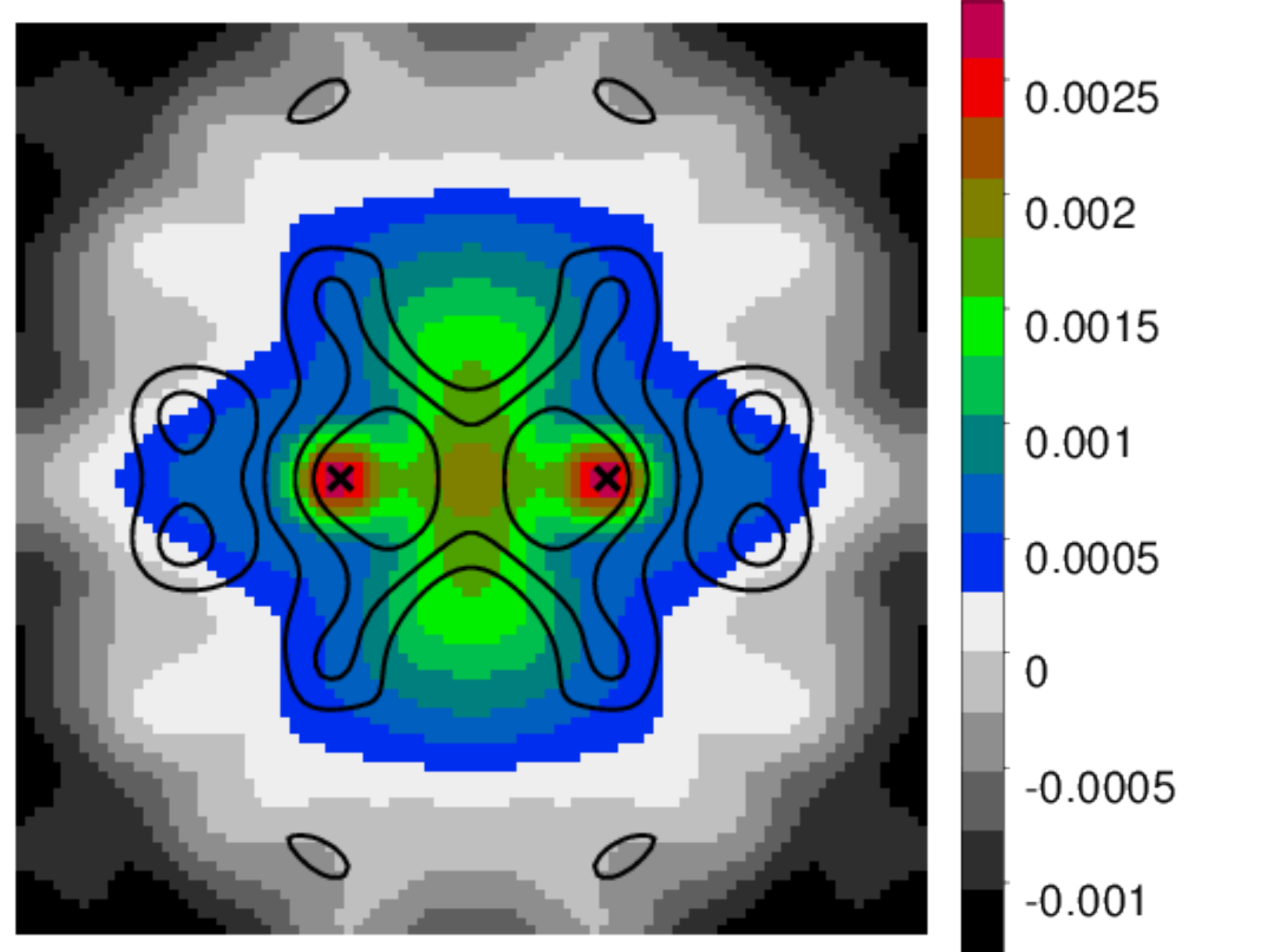}
  \end{center}
  \caption{\label{fig:bmodesall} The average of all pair-convergence
    maps in Fig. \ref{fig:comparison}, panels K, L, O, and P, and the
    corresponding B-mode maps by combining all shear stacks used for
    that figure. The E-mode is shown here as intensity scale, the
    B-mode is shown as overlay of iso-contours for the levels
    $2\times10^{-4}$, $4\times10^{-4}$, and $6\times10^{-4}$. }
\end{figure*}

As additional characterisation of the population of lens galaxies in
our analysis of the excess mass, we plot in
Fig. \ref{fig:completeness} their distribution of absolute rest-frame
magnitudes or colours for (photometric) redshifts $z_{\rm ph}$ between
$0.2$ and $0.6$. Moreover, the Figures \ref{fig:bmodes} and
\ref{fig:bmodesall} are maps of the B-mode signal of the excess mass
which are a diagnostic for systematic errors in the correlation
signal.

\end{document}